\documentclass[a4paper]{article}

\usepackage{graphicx}
\usepackage{dcolumn}
\usepackage{bm}
\usepackage{amssymb}

\title{Flow pattern transition accompanied with sudden growth of flow resistance
in two-dimensional curvilinear viscoelastic flows}

\author{Hiroki Yatou \\
Department of Physics, Graduate School of Science, \\
Kyoto University, Kyoto 606-8502, Japan
}
\date{\today}

\begin{document}

\maketitle

\begin{abstract}
We find three types of steady solutions and remarkable flow pattern transitions between them
in a two-dimensional wavy-walled channel for low to moderate Reynolds ($\mathrm{Re}$)
and Weissenberg ($\mathrm{Wi}$) numbers using direct numerical simulations with spectral element method.
The solutions are called "convective", "transition", and "elastic" in ascending order of $\mathrm{Wi}$.
In the convective region in the $\mathrm{Re}$-$\mathrm{Wi}$ parameter space,
the convective effect and the pressure gradient balance on average.
As $\mathrm{Wi}$ increases, the elastic effect becomes suddenly comparable and the first transition sets in.
Through the transition, a separation vortex disappears and a jet flow induced close to the wall by the viscoelasticity moves into the bulk;
The viscous drag significantly drops and the elastic wall friction rises sharply.
This transition is caused by an elastic force in the streamwise direction due to the competition of the convective and elastic effects.
In the transition region, the convective and elastic effects balance.
When the elastic effect dominates the convective effect, the second transition occurs but it is relatively moderate. 
The second one seems to be governed by so-called Weissenberg effect.
These transitions are not sensitive to driving forces.
By the scaling analysis, it is shown that the stress component is proportional to the Reynolds number
on the boundary of the first transition in the $\mathrm{Re}$-$\mathrm{Wi}$ space.
This scaling coincides well with the numerical result.
\end{abstract}

\section{Introduction}
\label{sec:introduction}

Nonlinearities due to viscoelasticity in fluid flows lead to a variety of non-trivial flows
such as rod-climbing, secondary flows, and tubeless siphon \cite{Bird1:1987}.
In polymeric fluids, even though the amount of the polymer additives is very small,
such a viscoelastic nonlinearity is effective.
For decades, the understanding of the formation mechanism of the non-trivial flows
due to nonlinear viscoelasticity has been of academic and industrial interest.

The nonlinearity originates from a viscoelastic stress augmented by the strain of flows and the feedback of the elastic stress to flows. 
The target of our study is a flow pattern formation due to the nonlinear coupling in curvilinear channel shear flows.
Flow pattern formations due to the nonlinear coupling are typically observed in rod climbing, i.e. Weissenberg effect \cite{Weissenberg:1947}.
The rod climbing is caused by an elastic stress working to the centripetal direction of curved streamlines (called "hoop stress") as follows.
When a rotating rod is inserted into viscoelastic fluids, the rotating shaft forms shear flows in the fluids.
Because the shear flows stretch and align polymer molecules along streamlines,
an elastic stress works to the inward direction of streamlines by the tension of elongated polymer molecules.
The fluid transferred to the inward direction finally climbs up along the rod.

The earlier studies have shown that the hoop stress is also the primary effect inducing elastic instabilities
in Taylor-Couette flows \cite{Larson:1990,Groisman:1998}.
In this system, perturbed radial flows cause additional hoop stress and the hoop stress then further intensifies the radial flows.
The elastic instabilities are followed by flow pattern formations including Taylor-Couette vortices, rotating standing waves,
disordered oscillations, and time-dependent or time-invariant solitary vortices between the cylinders with rotational speed of an inner cylinder
\cite{Groisman:1996,Groisman:1998,Baumert:1999,Thomas1:2006}.
Elastic instabilities in curvilinear shear flows are also observed
in Taylor-Dean flow \cite{Joo1:1992}, cone-and-plate flow \cite{McKinley:1991}, and lid-driven cavity flow \cite{Pakdel:1996}.

These instabilities occur even at very low or even zero Reynolds number $\mathrm{Re}\ll1$.
In this regime, elasticity-induced turbulence has also been observed in various experiments of shear flows
including Taylor-Couette flow, von Karman swirling flow, and curvilinear channel flow \cite{Groisman:2000,Groisman:2001,Groisman2:2004}.
The characteristics of the elastic turbulence are an increased flow resistance and efficient mixing at very low Reynolds number \cite{Burghelea1:2004}.
This mixing is very useful for industrial applications such as micro/nano devices \cite{Squires:2005}.

Our goal of the investigations is the understanding of the detailed mechanism of flow pattern formation, elastic instability, and elastic turbulence
in viscoelastic curvilinear channel flows as employed in the experiments of elastic turbulence \cite{Groisman:2001,Burghelea1:2004,Pathak:2004}.
Although the previous studies concerning elastic instabilities in curved flows have basically noticed a radial force due to the hoop stress,
in this study, we focus on the flow pattern transition caused by a streamwise elastic force,
that is an elastic force in the streamwise direction in curved streamlines.
The transitions are characterized by the disappearance of the separation vortex and the emergence of a jet flow with the sudden variation of wall frictions, and 
are caused by the increase of streamwise flows along curved streamlines due to the streamwise elastic force.
Note that the flow pattern transition due to the streamwise elastic force is effective only when an inertial effect is also effective at moderate Reynolds number
unlike the earlier elastic instabilities for $\mathrm{Re}\ll1$.
This indicates that the competition of the convective and elastic effects is essential for the flow pattern transition in the curvilinear channel shear flows.

In order to investigate a detailed mechanism of flow pattern formation,
direct numerical simulation is an effective approach on such complex flows.
Nevertheless, earlier direct numerical simulations in the context of elastic instabilities
have not treated such complex flows but treated more simple flows including two-dimensional Kolmogorov and four-mill flows
driven by external forcing in doubly-periodic boundary conditions which have no boundary walls \cite{Thomas1:2006,Berti:2008,Thomases:2009}.
Therefore, we perform direct numerical simulations of periodic shear flows in two-dimensional wavy-walled channels as sketched in Fig.\ \ref{fig:geometry}
as a model of curvilinear channel shear flows with high-accuracy spectral element method.
The periodic wavy-walled geometry is canonical for flows with open and curved streamlines \cite{Sadanandan:2004}
and has actually also attracted much attention for industrial applications such as compact heat exchangers \cite{Asako:1988,Cho:1998}.

In this study, we investigate base laminar steady solutions for low to moderate Reynolds and Weissenberg numbers.
Note that Newtonian flows in this system is linearly unstable at moderate Reynolds number \cite{Cho:1998}
unlike two-dimensional Taylor-Couette and Taylor-Dean flows of viscoelastic fluids \cite{Joo2:1992,Renardy:1986}.
This linear instability may show that viscoelastic fluid flows are also linearly unstable in wavy-walled channels.
We expect that the study of the base solutions leads to future elastic instability investigations.

This paper is organized in the following way.
Section \ref{sec:vequations} provides governing equations and several non-dimensional parameters. 
A numerical method, boundary conditions and a flow geometry are described in Sec.\ \ref{sec:vnumerical}.
The validation of our computations is done and parameter values used in this study are summarized in Sec.\ \ref{sec:vnumerical}.
Section \ref{sec:vresults} shows the main results of our study.
First, the steady solutions is classified into three groups according to the flow patterns.
Second, the variation of a viscous wall friction, a flow rate, and an elastic wall friction with Weissenberg number is described.
Third, we describe the mechanism of the flow pattern transition and the variation of the above flow characteristics
by utilizing the relationship among the norms of the terms in the Navier-Stokes equation and among local forces.
Fourth, we perform scaling analyses in order to confirm the relationship of the norms.
Fifth, the flow pattern formation in Poiseuille-type flows driven by constant body force are also shown.
Section \ref{sec:vdiscussion} is devoted to discussion and concluding remarks.

\section{Basic equations}
\label{sec:vequations}

The equations governing viscoelastic fluid flows consist of an equation of continuity,
a momentum equation and a constitutive equation describing the temporal evolution of a rheological state.
We employ the incompressible Navier-Stokes equation for the momentum equation and the FENE-P model for the constitutive equation.
The governing equations are written in the following dimensionless form:
\begin{equation}
 \nabla\cdot{\bm u} = 0,
 \label{eq:ncontinuity}
\end{equation}
\begin{equation}
\mathrm{Re}\left[\frac{\partial{\bm u}}{\partial t}+({\bm u}\cdot\nabla){\bm u}\right]
= -\nabla p + \beta\nabla^2{\bm u}+\nabla\cdot{\bm\tau},
 \label{eq:nnavierstokes}
\end{equation}
\begin{eqnarray}
\mathrm{Wi}\left[\frac{\partial {\bm C}}{\partial t}+({\bm u}\cdot\nabla){\bm C}
-(\nabla{\bm u})^{T}\cdot{\bm C}-{\bm C}\cdot(\nabla{\bm u})-\kappa\Delta{\bm C}\right] \nonumber \\
+ (f(r){\bm C}-{\bm I}) = 0,
 \label{eq:nfenep}
\end{eqnarray}
\begin{equation}
 {\bm\tau}=\frac{1-\beta}{\mathrm{Wi}}(f(r){\bm C}-{\bm I}),
 \label{eq:nstress}
\end{equation}
\begin{equation}
 f(r) = \frac{K}{1-r^2/L^2}, \quad K=1-3/L^2,
 \label{eq:peterlin}
\end{equation}
where $\bm u$ is the velocity, $p$ is the pressure, and $\bm\tau$ is the extra-stress tensor.
$\bm C$ is the conformation tensor and $L$ is the maximum polymer length.

The equations include the following non-dimensional parameters:
the Reynolds number, the Weissenberg number, and the viscosity ratio parameter.
The Reynolds number is the ratio of an inertial force to a viscous force defined in the following way:
\begin{equation}
\mathrm{Re}= \rho U_0 L_0/\eta,
\end{equation}
where $\rho$ is the fluid density, $L_0$ is the typical length of flow geometry, and $U_0$ is the typical flow speed.
$\eta=\eta_s+\eta_p$ is the total viscosity defined as the sum of the solvent viscosity $\eta_s$ and the polymer viscosity $\eta_p$.
The Weissenberg number is the ratio of a polymer relaxation time and a typical flow time:
\begin{equation}
\mathrm{Wi} = \lambda U_0/L_0,
\label{eq:weissen}
\end{equation}
where $\lambda$ is the Zimm relaxation time.
The viscosity ratio parameter is the ratio of the solvent viscosity and the total viscosity:
\begin{equation}
\beta = \eta_s / \eta.
\end{equation}

From the microscopic viewpoint,
the FENE-P model is derived from a kinetic theory \cite{Bird:1980} and the pre-averaging closure approximation introduced by Peterlin \cite{Peterlin:1966}
concerning elastic polymer dumbbells consisting of two beads and a spring connecting them.
The conformation tensor $\bm C$ represents a measure of the second-order moment of the end-to-end distance vector of polymer dumbbells:
\begin{equation}
 C_{ij} = \langle r_ir_j\rangle,
\end{equation}
where $\bm r$ is the relative difference of two beads positions.
The function $f(r)$ in Eq.\ (\ref{eq:peterlin}) is so-called Peterlin function,
which limits the length of polymers to be lower than the maximum polymer length $L$.
When $L\to\infty$, the FENE-P model coincides with the Oldroyd-B model,
which is derived from a kinetic theory on linear Hookean dumbbells.

Because the FENE-P model reproduces the flow properties qualitatively similar to those observed in experiments, for example "shear thinning"
\cite{Bird2:1987},
the FENE-P model has been frequently employed in many studies on drag reduction \cite{Sureshkumar:1997,Dimitropoulos:1998}
, elastic instability \cite{Kumar:2001,Thomas1:2006}, and the other general viscoelastic fluid flows.
We hence adopt the FENE-P model as a constitutive equation.

\subsection{Additional diffusive term}

In order to avoid numerical instabilities,
we add an artificial diffusive term $\kappa\nabla^2{\bm C}$ in the constitutive equation (\ref{eq:nfenep}),
where $\kappa$ is the artificial constitutive diffusivity coefficient.
The artificial diffusive term added to the constitutive equation physically originates from the Brownian motion
of the center-of-mass of elastic dumbbells across streamlines \cite{El-Kareh:1989}.
Sureshkumar \cite{Sureshkumar1:1995} investigated the influence of the addition of the artificial diffusive term
on flow stability in detail.
They concluded that sufficiently small artificial diffusivity coefficient ($\kappa\leq10^{-3}$) 
does not affect the critical eigenmodes of the viscoelastic Orr-Sommerfeld problem though
the artificial diffusivity coefficient needed for numerical stabilization is much larger than the values physically obtained in the earlier study \cite{El-Kareh:1989}.
This implies that sufficiently small artificial diffusivity coefficient does not qualitatively change flow behavior of viscoelastic fluids
from the viewpoint of numerical instabilities.
This additional diffusivity hence has been widely used in the calculations of viscoelastic fluid flows governed with 
the Oldroyd-B, FENE-P, and other models \cite{Thomases:2009,Atalik:2002}.

\section{Numerical method}
\label{sec:vnumerical}

\subsection{Spectral element methods}

We adopt spectral element method (hereinafter referred to as SEM)
to numerically solve the governing equations in a complex geometry such as wavy-walled channels with high accuracy
\cite{Patera:1984,Owens:2002,Canuto:2007}.
In SEM, a computed domain is subdivided into several spectral elements.
The values of variables and the derivatives of the variables are interpolated through the values on grid points
with Lagrange polynomial interpolation functions in the elements \cite{Boyd:2001}.
Numerical errors between numerical solutions and accurate solutions are reduced in two ways in SEM.
First, the increase of the polynomial degree of interpolation functions
results in exponential convergence of numerical accuracy for sufficiently smooth solutions \cite{Maday:1989}.
Second, the increase of the number of elements leads to an algebraic convergence
similar to that of finite element calculations.
SEM thereby enables the calculations with flexible boundary conditions and high numerical accuracy
which are advantages of finite element methods and spectral methods, respectively.
Because the local mesh refinement is required for viscoelastic calculations due to an elastic stress enhancement near walls,
we provide more grid points near walls as shown in Fig.\ \ref{fig:geometry}(b).

The Galerkin formulation of the primitive equations (\ref{eq:ncontinuity})-(\ref{eq:nfenep}) is used for spatial discretization in SEM.
The Galerkin formulation in a two-dimensional domain $\Omega$ is written as follows:
\begin{equation}
 \int_{\Omega}(\nabla\cdot{\bm u})q d\Omega = 0,
 \label{eq:vgalerkincont}
\end{equation}
\begin{eqnarray}
\int_{\Omega}\frac{\partial{\bm u}}{\partial t}\cdot{\bm w}d\Omega - 
\int_{\Omega}p({\bm I}:\nabla{\bm w})d\Omega + 
\mathrm{Re}\int_{\Omega}[({\bm u}\cdot\nabla){\bm u}]\cdot{\bm w}d\Omega \nonumber \\
 + \beta\int_{\Omega}(\nabla{\bm u}:\nabla{\bm w})d\Omega + \int_{\Omega}{\bm\tau}:(\nabla{\bm w})d\Omega = 0, \nonumber \\
 \label{eq:vgalerkinnavier}
\end{eqnarray}
\begin{eqnarray}
\mathrm{Wi}\int_{\Omega}\left[\frac{\partial{\bm C}}{\partial t} + ({\bm u}\cdot\nabla){\bm C}
- ((\nabla{\bm u})^T\cdot{\bm C}) - {\bm C}\cdot(\nabla{\bm u})\right]:{\bm S}d\Omega \nonumber \\
+ \mathrm{Wi}\int_{\Omega}(\kappa\nabla{\bm C}):(\nabla{\bm S})d\Omega + \int_{\Omega}(f(r){\bm C}-{\bm I}):{\bm S}d\Omega = 0 \nonumber \\
 \label{eq:vgalerkinconst}
\end{eqnarray}
where the tensor product is defined for two arbitrary tensors $\bm P$ and $\bm Q$ as:
\begin{equation}
{\bm P}:{\bm Q} = P_{ij}Q_{ij},\quad 1\leqslant i,j \leqslant 2
\end{equation}
and the notation $(\nabla{\bm u})_{ij} = \partial u_j/\partial x_j$ is adopted.
The pressure $p$ and the test function $q$ belong to the functional space $Q\subset L^2(\Omega)$.
The velocity $\bm u$ and the test function $\bm w$ belong to the space $W\subset H^1(\Omega)^2$.
The conformation tensor $\bm C$ and the test function $\bm S$ belong to $\Sigma \subset H^1(\Omega)^{2\times 2}$.
Here $L^2(\Omega)$ and $H^1(\Omega)$ represent the space of square-integrable functions
and the space of derivative functions whose first order derivatives are square-integrable on $\Omega$.
The integral of these equations is performed with Lagrange interpolation functions and quadrature rules.
Finally, nonlinear algebraic equations described with operator matrices and vectors of variables are derived.
The linear terms are discretized by a second-order backward difference (BDF2) scheme and 
the nonlinear terms are discretized by a second-order extrapolation (EX2) scheme.
We adopt a decoupled approach to solve time-splitting equations \cite{Fietier2:2003}.

\subsection{Boundary condition}

No-slip boundary conditions are imposed at the upper and bottom walls for the velocity.
The boundary integral, which stems from the partial integral
of the artificial diffusive term of the constitutive equation, assumes to be zero for numerical convenience.
A periodic boundary condition is imposed on the velocity and the conformation tensor on inflow and outflow boundaries.
No boundary condition is imposed on the pressure because the pressure has no grid points on the boundaries.

We mainly investigate Couette-type flows in which the upper wall is at rest and constant speed is given on the bottom wall.
The wall speed of the bottom wall is normalized to unity along tangential directions $t$
denoted in Fig.\ \ref{fig:geometry}(a) at every point on the bottom wall.
We also investigate Poiseuille-type flows in which both the upper and bottom walls are at rest.

\subsection{Parameters}

Here, we summarize the parameters used in this study:
(1) fluid and rheological parameters are the Reynolds number $\mathrm{Re}$, the Weissenberg number $\mathrm{Wi}$,
the viscosity ratio parameter $\beta$, and the maximum polymer length $L$,
(2) geometric parameters are the wall amplitude $A$ and the periodic length $L_x$ defined at Sec.\ \ref{sec:geometry},
(3) the artificial diffusivity coefficient $\kappa$.
We focus on the dependence of flow behavior on the Reynolds number, the Weissenberg number,
and the curvature of streamlines controlled by the wall amplitude $A$.
The other parameters $\beta$, $L$, $\kappa$, and $L_x$ are fixed to the values shown in Tab.\ \ref{tab:vparameters}.
We set the maximum polymer length $L$ to the sufficiently large value $L^2=1.0\times10^4$
in order to capture the influence of a large elastic stress on fluid flows.
The values of the fixed parameters $\beta$, $L$, and $\kappa$ have been frequently used in the previous numerical studies
of viscoelastic fluid flows \cite{Thomases:2009,Thomas1:2006,Sureshkumar1:1995,Atalik:2002}.
Our several test runs show that these parameter values do not qualitatively influence the flow behavior obtained in this study.

\begin{table}
 \begin{center}
   \begin{tabular}{ccccccc} \hline\hline
     $N_{el}$  &  $N_v$ & $L_x$  & $\beta$  & $L^2$           & $\kappa$           \\ \hline
      $192$    &  $6$   &  $\pi$ & $0.5$    & $1.0\times10^4$ & $1.0\times10^{-3}$ \\ \hline\hline
   \end{tabular}
 \end{center}
 \caption{Grid point numbers and parameter values fixed in this study. 
 $N_{el}$ is the number of elements and $N_v$ is the order of interpolation polynomials of the velocity. }
 \label{tab:vparameters}
\end{table}

\subsection{Validation of our computations}

Validation of our numerical code is done on plane-Couette flows.
We perform the calculations at the low Weissenberg number $\mathrm{Wi}=1.0$ with zero artificial diffusivity coefficient $\kappa=0$
and at the high Weissenberg number $\mathrm{Wi}=20.0$ with the finite artificial diffusivity coefficient $\kappa=1.0\times10^{-3}$.

We compare our numerical solutions with exact steady solutions of the FENE-P model.
The exact solutions of plane-Couette flows of the FENE-P model are derived as the roots of the following equations (\ref{eq:check1}) and (\ref{eq:check2})
obtained from the original equations under the condition that the derivatives of all variables with respect to $x$ are zero,
$p=\mathrm{const}$, $u_x=1-y$, and $u_y=0$:
\begin{equation}
2\mathrm{Wi}^2C_{yy}^3+(L^2K+2)C_{yy}-L-2 = 0,
\label{eq:check1}
\end{equation}
\begin{equation}
C_{xy} + \mathrm{Wi}C_{yy}^2 = 0, \quad C_{xx} - L^2 + (L^2K+1)C_{yy} = 0.
\label{eq:check2}
\end{equation}
The conformation tensor $\bm C$ of the exact solutions is uniform in the whole domain
and the extra-stress component $\tau_{yy}$ is zero everywhere.
We calculate the extra-stress component $\tau_{yy}$ for Validation of our computations.

Table.\ \ref{tab:validation} shows the absolute values of the extra-stress component $\tau_{yy}$ with several parameter values.
The difference between our results and the exact solutions is within $10^{-6}$ for the wide range of parameter values.
Moreover, numerical accuracy is improved with mesh refinements.

\begin{table}
 \begin{center}
   \begin{tabular}{cccccc} \hline\hline
     $N_{el}$ &  $N_v$ & $\mathrm{Wi}$ & $L^2$           & $\kappa$          &  $|\tau_{yy}|$          \\ \hline
     $48$     &  $4$   &   $1.0$       & $10.0$          & $0.0$             &  $3.487\times10^{-7}$   \\ 
    $192$     &  $6$   &   $1.0$       & $10.0$          & $0.0$             &  $4.200\times10^{-9}$   \\ 
     $48$     &  $4$   &   $1.0$       & $1.0\times10^4$ & $0.0$             &  $2.999\times10^{-7}$   \\
    $192$     &  $6$   &   $1.0$       & $1.0\times10^4$ & $0.0$             &  $1.583\times10^{-9}$   \\
    $48$      &  $4$   &   $20.0$      & $1.0\times10^4$ & $1.0\times10^3$   &  $5.609\times10^{-7}$   \\ 
   $192$      &  $6$   &   $20.0$      & $1.0\times10^4$ & $1.0\times10^3$   &  $1.050\times10^{-8}$   \\ \hline\hline
   \end{tabular}
 \end{center}
 \caption{Absolute value of the extra-stress component $\tau_{yy}$ with several parameter values for validation of our numerical code.  }
 \label{tab:validation}
\end{table}

\subsection{flow geometry}
\label{sec:geometry}

\begin{figure}
  \centerline{\includegraphics[height=160pt]{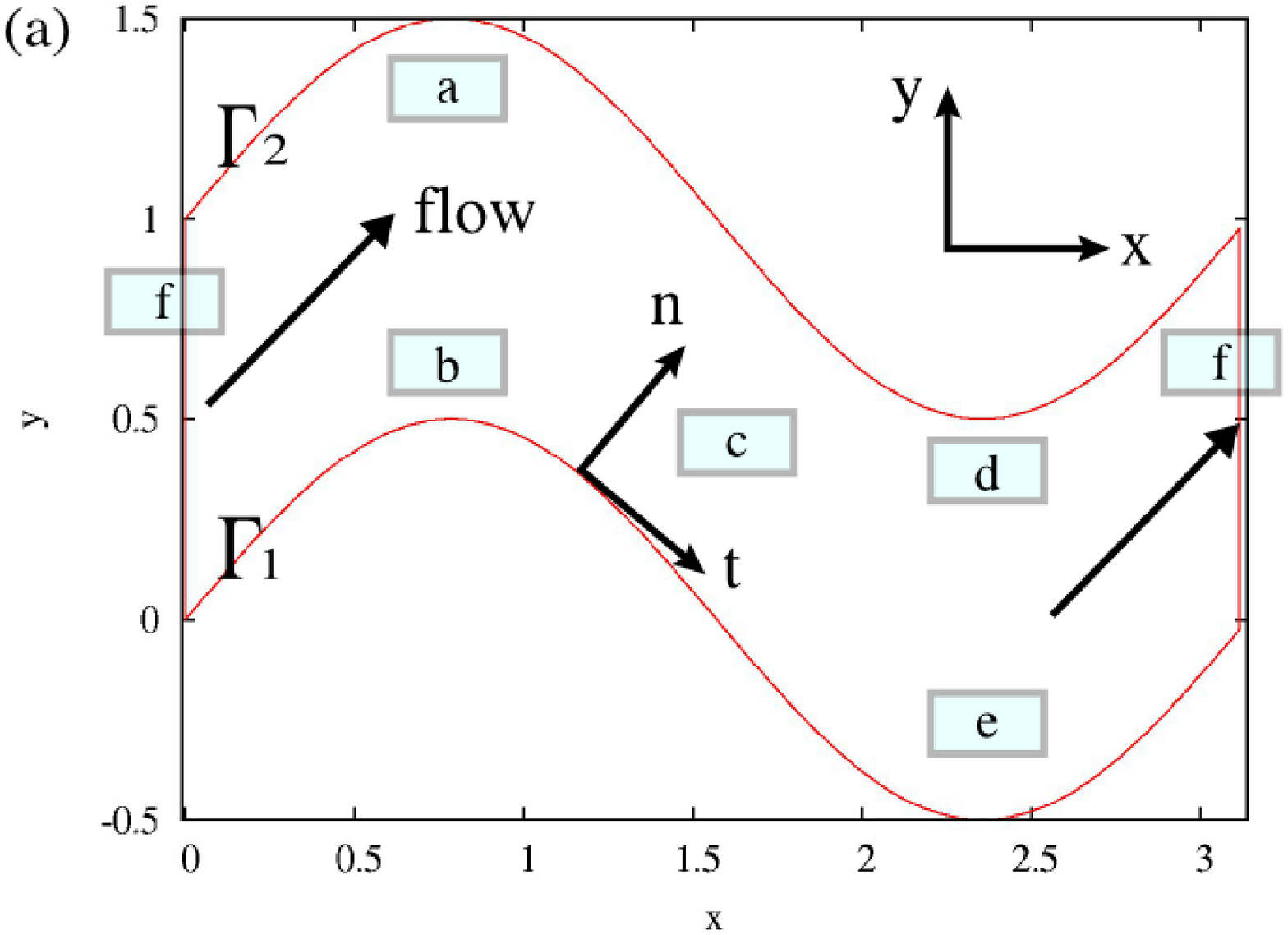}
  \includegraphics[height=160pt]{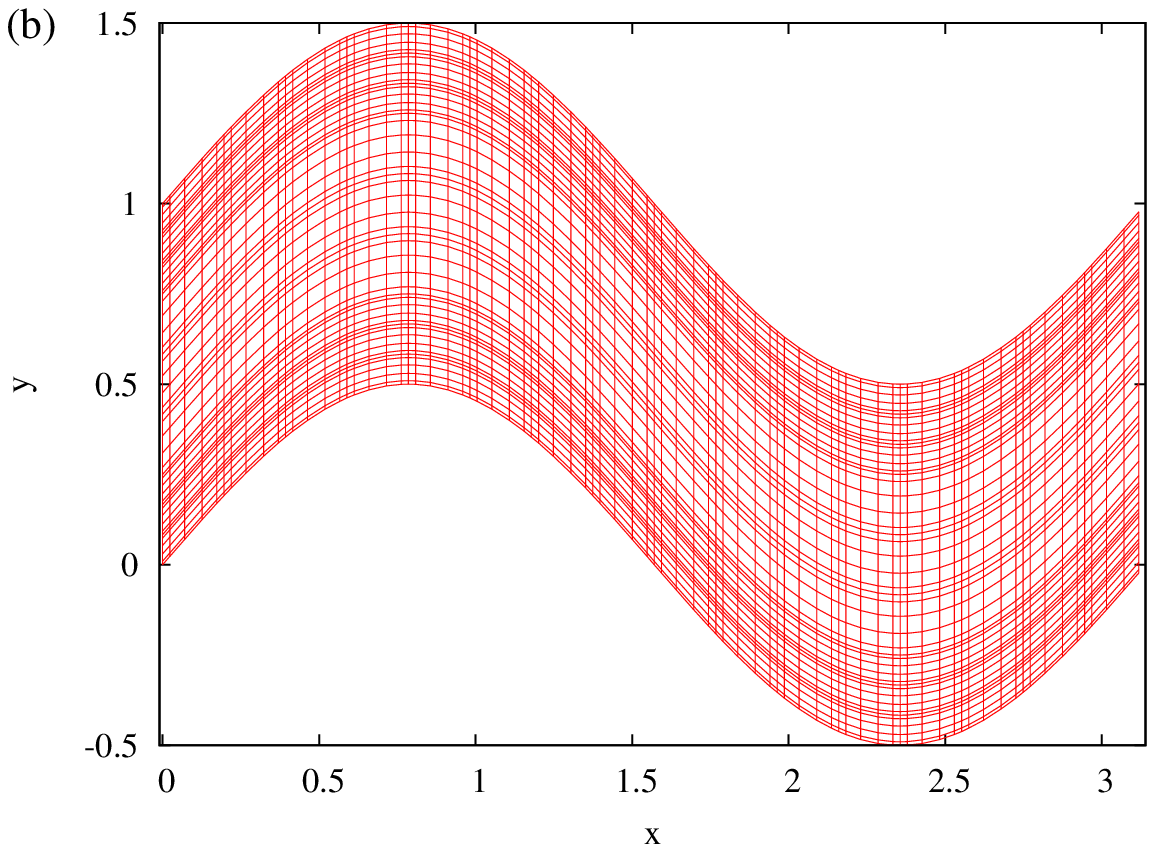} }
  \caption{(Color online) (a) Flow domain with symbols and expressions.
  The regions "a" and "e" are the outward regions, "b" and "d" are the inward regions, and "c" and "f" are the contraction regions.
  The symbols "t" and "n" denote a tangential and a normal directions to the boundary wall, respectively.
  (b) Computational mesh used in this study. }
  \label{fig:geometry}
\end{figure}

The governing equations are solved in the domain sketched in Fig.~\ref{fig:geometry}(a).
Two sinusoidal curves are boundary walls.
The coordinates of the upper and bottom walls are defined as follows:
\begin{eqnarray}
\mbox{(upper):} & y & = 1 + A\sin(2\pi x/L_x), \nonumber \\
\mbox{(bottom):} & y & = A\sin(2\pi x/L_x), \nonumber
\end{eqnarray}
where $A$ is the amplitude of wall corrugations and $L_x$ is the length of the periodic channel in the direction of $x$.
The width of the channel is normalized to unity.
We denote this channel "sinuous channel" as called by Cho \cite{Cho:1998}.
We call the regions denoted as "a" and "e" in Fig.\ \ref{fig:geometry}(a) "outward region"
because these regions are located in the outward position of curved flows parallel to the wall.
We similarly call the regions denoted as "b" and "d" in Fig.\ \ref{fig:geometry}(a) "inward region".
The regions denoted as "c" and "f" in Fig.\ \ref{fig:geometry}(a) are expressed "contraction region" in this paper.
The symbols "t" and "n" denote a tangential and a normal directions to the wall, respectively.

We adopt the sinuous channel in this study for the following reasons.
First, the channel has mixed-kinematics with both shear and extensional flows.
The elastic stress due to polymer stretching in these flows strongly influences flow behavior.
Second, we can easily control the magnitude of shear and extension by changing the wall amplitude $A$ and the periodic length $L_x$ of the channel.
Third, streamlines are highly curved in the channel.
The curvature of streamlines locally varies and the sign of the curvature is even reversed along the streamlines
unlike two-dimensional Taylor-Couette flows in which the sign of the curvature is not changed along streamlines.
The spatial variation of the curvature leads to the change of the magnitude and the direction of an elastic hoop stress referred at Sec.\ \ref{sec:introduction}.
The spatial variation of the hoop stress plays a crucial role in the formation of flow patterns as described below.

Despite the importance of the sinuous channel,
most of the earlier works of viscoelastic fluid flows in wavy-walled channels
have considered corrugated (contraction/expansion) channel flows in which the phases of sinusoidal curves
of an upper and a bottom walls are shifted in a half wavelength \cite{Sureshkumar:2001,Sadanandan:2004,Kemenade:1994,Arora:2002}.
The studies of the sinuous channel flows are limited to those of Newtonian fluid flows \cite{Asako:1988,Cho:1998}.
However, the corrugated channel is not appropriate for our study focusing on the curvilinear channel flows
because the streamlines of a bulk flow in the corrugated channel are more weakly curved than those in the sinuous channel.

\section{Results}
\label{sec:vresults}

\subsection{Flow pattern transition and classification of steady solutions}
\label{sec:classification}

We perform direct numerical simulations for the wide range of Weissenberg number
$0\leq\mathrm{Wi}\leq20$ and Reynolds number $0\leq\mathrm{Re}\leq1000$.
We obtain steady solutions in these regions.

\begin{figure}
  \centerline{\includegraphics[height=200pt]{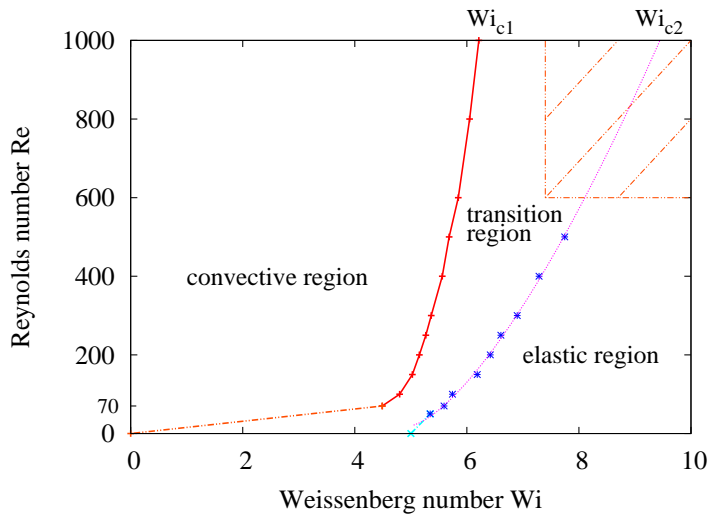}}
  \caption{(Color online) Phase diagram of the base steady solutions in the plane of $\mathrm{Wi}$ and $\mathrm{Re}$.
  The left and right curves denote the first and second critical Weissenberg numbers,
  $\mathrm{Wi}_{\mathrm{c}1}$ and $\mathrm{Wi}_{\mathrm{c}2}$, defined at Sec.\ \ref{sec:balance}, respectively. 
  The Reynolds number of the edge point on the left curve $\mathrm{Re}=70$ is determined through the scaling analysis at Sec.\ \ref{sec:scaling}.
  The curve of $\mathrm{Wi}_{\mathrm{c}2}$ is extrapolated to a large-$\mathrm{Re}$ region using a fitting function. 
  The value of $\mathrm{Wi}_{\mathrm{c}2}$ at $\mathrm{Re}=0$ is the Weissenberg number at which the separation vortex emerges.  
  The region with diagonal lines represents that in which our calculations suffer numerical instabilities.  }
  \label{fig:phase}
\end{figure}

The steady solutions are classified into three groups.
The three groups correspond to three regions in the $\mathrm{Wi}$-$\mathrm{Re}$ parameter space as shown in Fig.\ \ref{fig:phase}.
We call the three regions "convective", "transition", and "elastic" regions.
The two Weissenberg numbers separating adjacent regions is called first
and second critical Weissenberg numbers, $\mathrm{Wi}_{\mathrm{c}1}$ and $\mathrm{Wi}_{\mathrm{c}2}$, defined at Sec.\ \ref{sec:balance}.
$\mathrm{Wi}_{\mathrm{c}2}$ is extrapolated into a large-$\mathrm{Re}$ region
using a fitting function because our numerical calculations suffer numerical instabilities at large $\mathrm{Wi}$ and $\mathrm{Re}$
shown in the region with diagonal lines in Fig.\ \ref{fig:phase}.

\begin{figure}
  \centerline{\includegraphics[height=500pt]{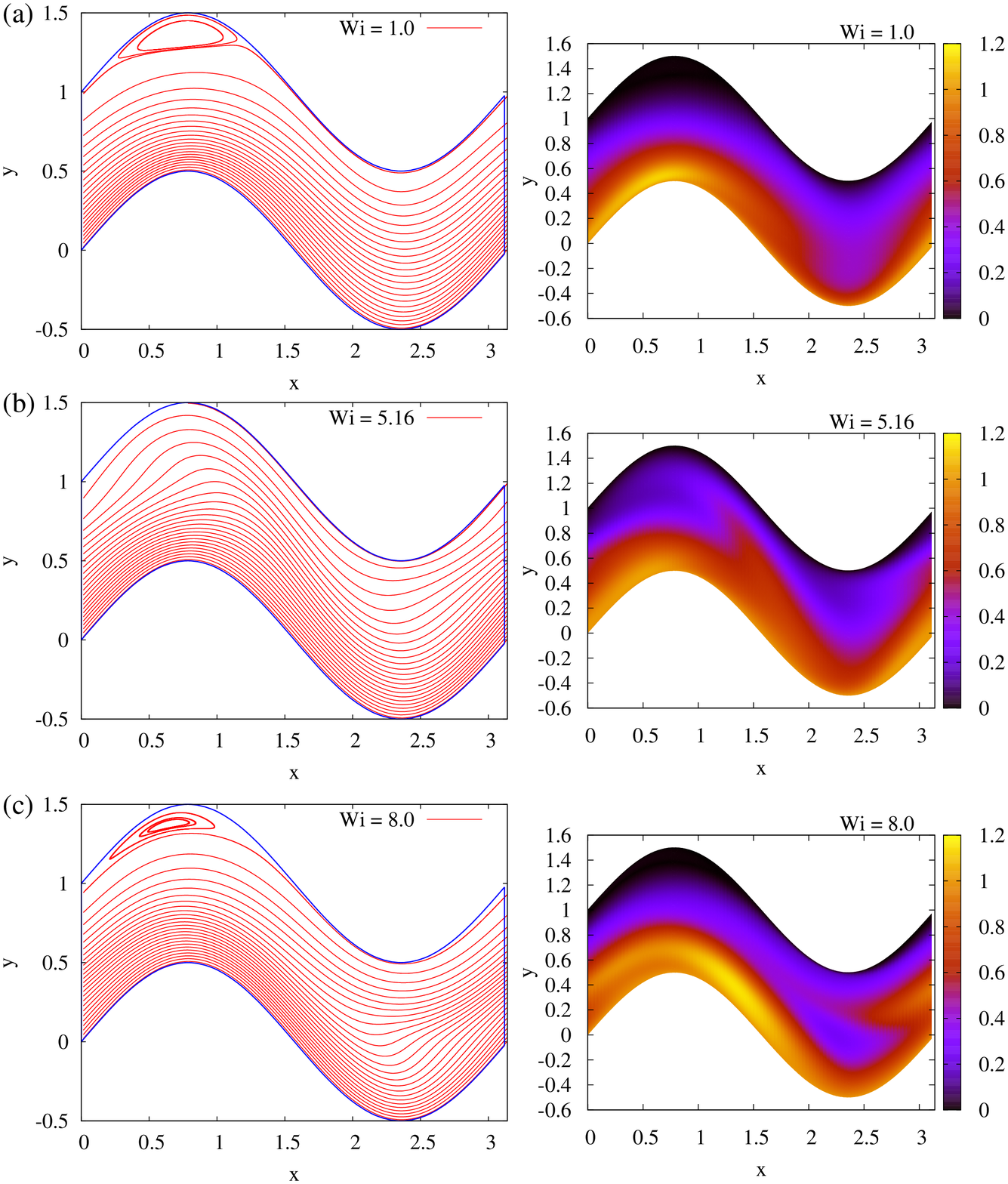} }
  \caption{(Color online) Streamlines (left) and flow-speed profiles (right) of the steady solutions in (a) the convective region $\mathrm{Wi}=1.0$,
   (b) the transition region $\mathrm{Wi}=5.16$, and (c) the elastic region $\mathrm{Wi}=8.0$ at $\mathrm{Re}=200$.
   Streamlines are drawn using passively-advected particles.
   The intervals of streamlines in bulk flows are determined to be inversely proportional to the flow speed at the position.
   The separation vortices are drawn independently of the bulk flows.  }
  \label{fig:streamline200}
\end{figure}

\begin{figure}
  \centerline{\includegraphics[height=500pt]{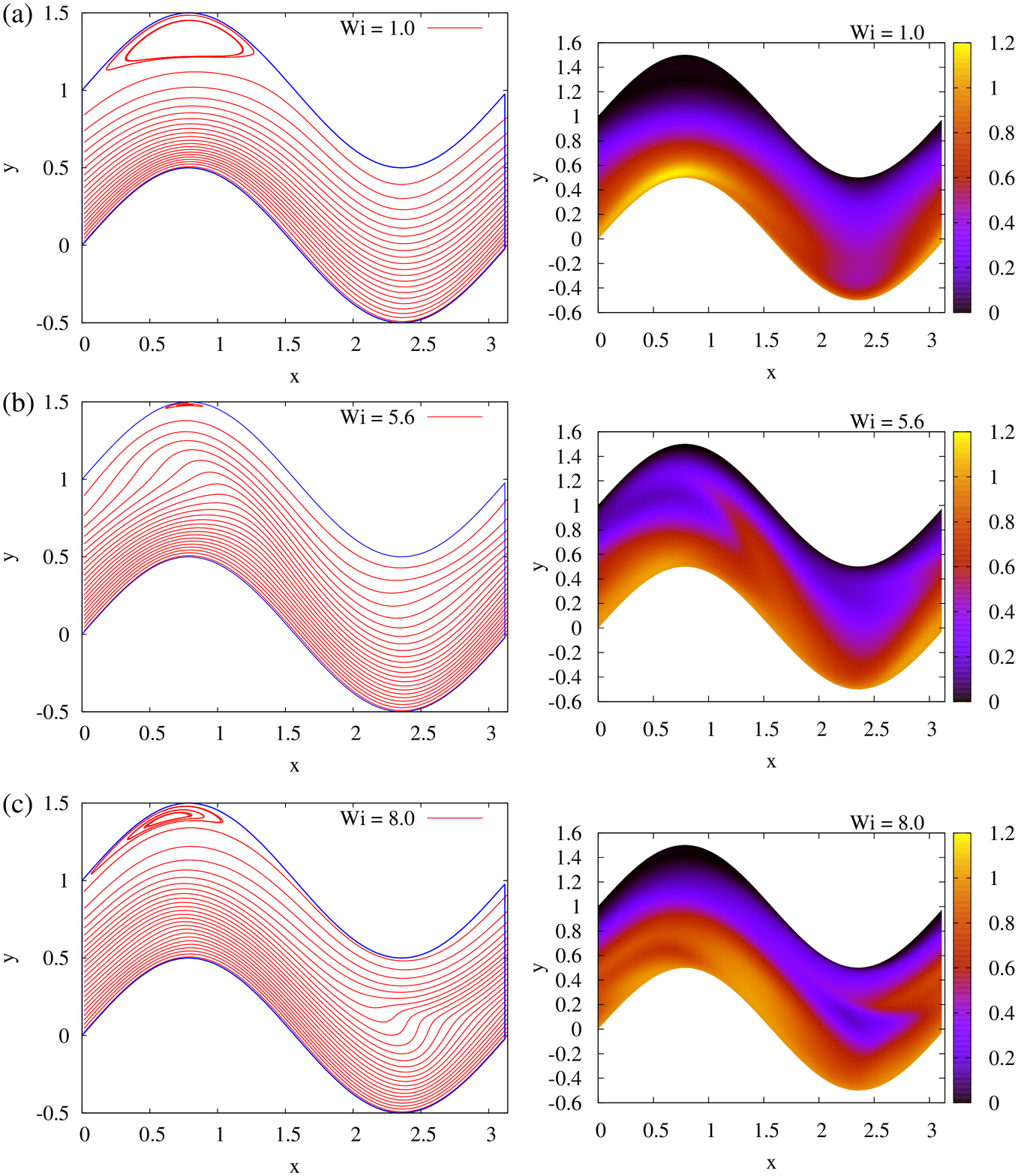} }
  \caption{(Color online) Streamlines (left) and flow-speed profiles (right) of the steady solutions in (a) the convective region $\mathrm{Wi}=1.0$,
         (b) the transition region $\mathrm{Wi}=5.6$, and (c) the elastic region $\mathrm{Wi}=8.0$ at $\mathrm{Re}=400$.
         The streamlines are drawn with the same method in Fig.\ \ref{fig:streamline200}. }
  \label{fig:streamline400}
\end{figure}

The difference of the steady solutions in the three groups
typically appears in the flow patterns of the steady solutions.
Figures \ref{fig:streamline200} and \ref{fig:streamline400} show streamlines and flow-speed profiles of the steady solutions
in the convective region $\mathrm{Wi}=1.0$, the transition region $\mathrm{Wi}=5.16$,
and the elastic region $\mathrm{Wi}=8.0$ at the moderate Reynolds numbers $\mathrm{Re}=200$ and $400$.
Figures\ \ref{fig:streamline200}(a) and \ref{fig:streamline400}(a) show that, in the convective region,
the flow field consists of a bulk shear flow and a vortex in the outward region (a)
(see the definition of the outward region (a) in Fig.\ \ref{fig:geometry}(a)).
We call the vortex "separation vortex" because the vortex is detached from the bulk flow in the channel.
The separation vortex has a symmetrical shape at low $\mathrm{Wi}$, while
the vortex becomes smaller and the symmetry of the vortex becomes gradually broken with $\mathrm{Wi}$.
The separation vortex is finally vanished or becomes very small near the upper wall
at $\mathrm{Wi}_{\mathrm{c}1}$ as shown in Fig.\ \ref{fig:streamline200}(b).
As discussed in detail, the depression of the separation vortex
is just caused by the increase of flow due to the streamwise elastic force in the outward region (a).
The separation vortex reappears and subsequently grows with $\mathrm{Wi}$ as shown in Fig.\ \ref{fig:streamline200}(c).

\begin{figure}
  \centerline{\includegraphics[height=500pt]{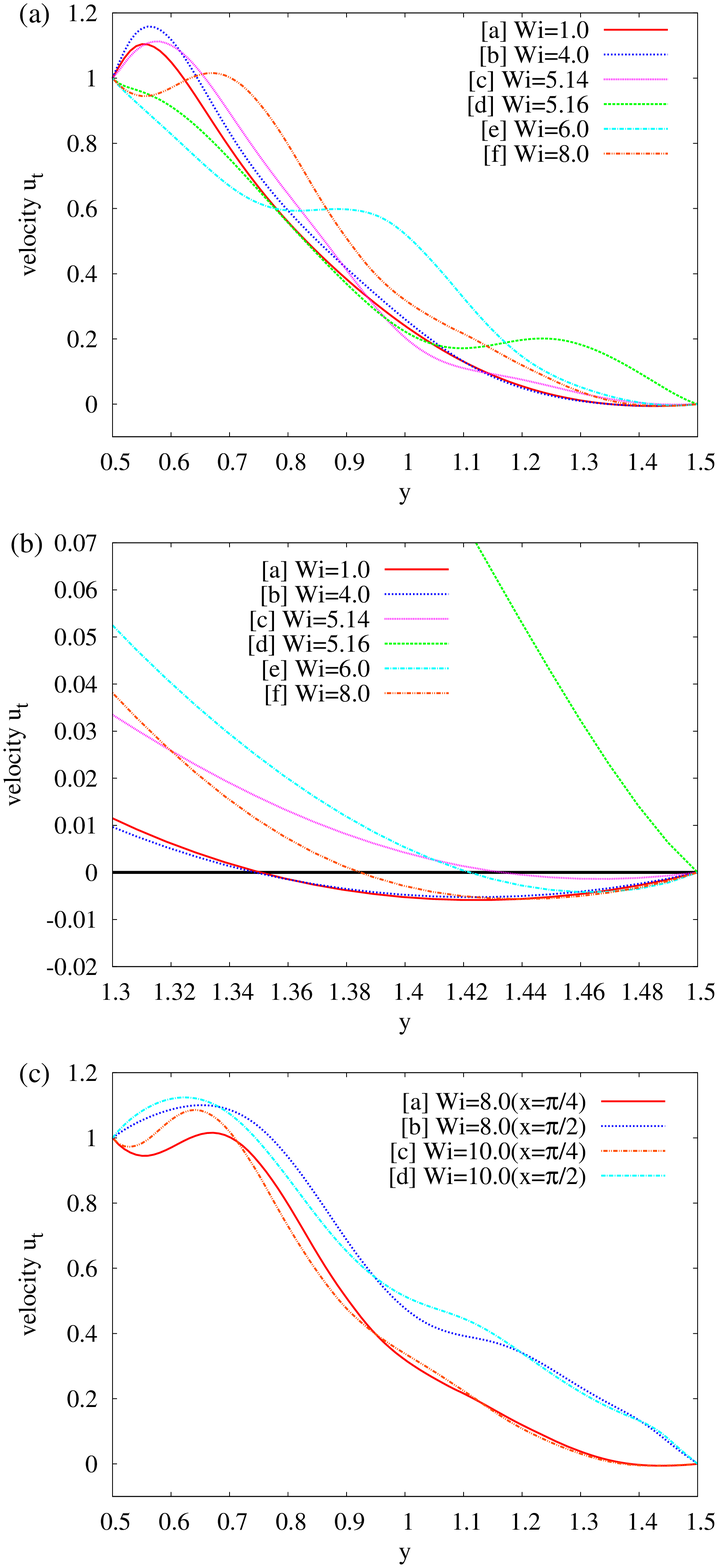}}
  \caption{(Color online) (a) Tangential velocities $u_t$ in the vertical section at $x=\pi/4$ in the convective region
  [a] $\mathrm{Wi}=1.0$ [b] $\mathrm{Wi}=4.0$, [c] $\mathrm{Wi}=5.1$, in the transition region [d] $\mathrm{Wi}=5.16$, 
  [e] $\mathrm{Wi}=5.5$, and in the elastic region [f] $\mathrm{Wi}=8.0$ at $\mathrm{Re}=200$.
  (b) Tangential velocities $u_t$ in the vertical section at $x=\pi/4$ magnified in $1.3\leq y\leq1.5$. 
  (c) Tangential velocities $u_t$ in the elastic region [a],[b] $\mathrm{Wi}=8.0$ and [c],[d] $\mathrm{Wi}=10.0$
  in the vertical section at [a],[c] $x=\pi/4$ and [b],[d] $x=\pi/2$. }
  \label{fig:flowsect}
\end{figure}

\begin{figure}
  \centerline{\includegraphics[height=170pt]{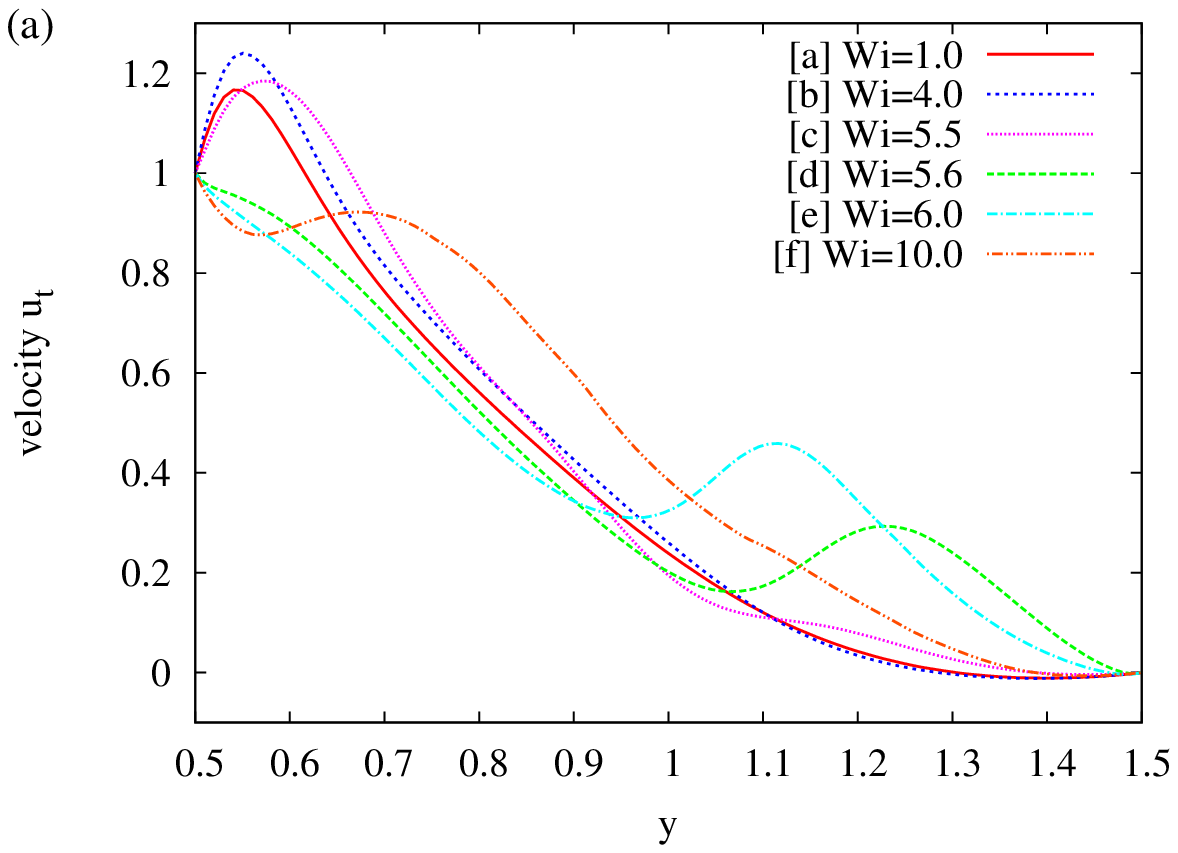}
  \includegraphics[height=170pt]{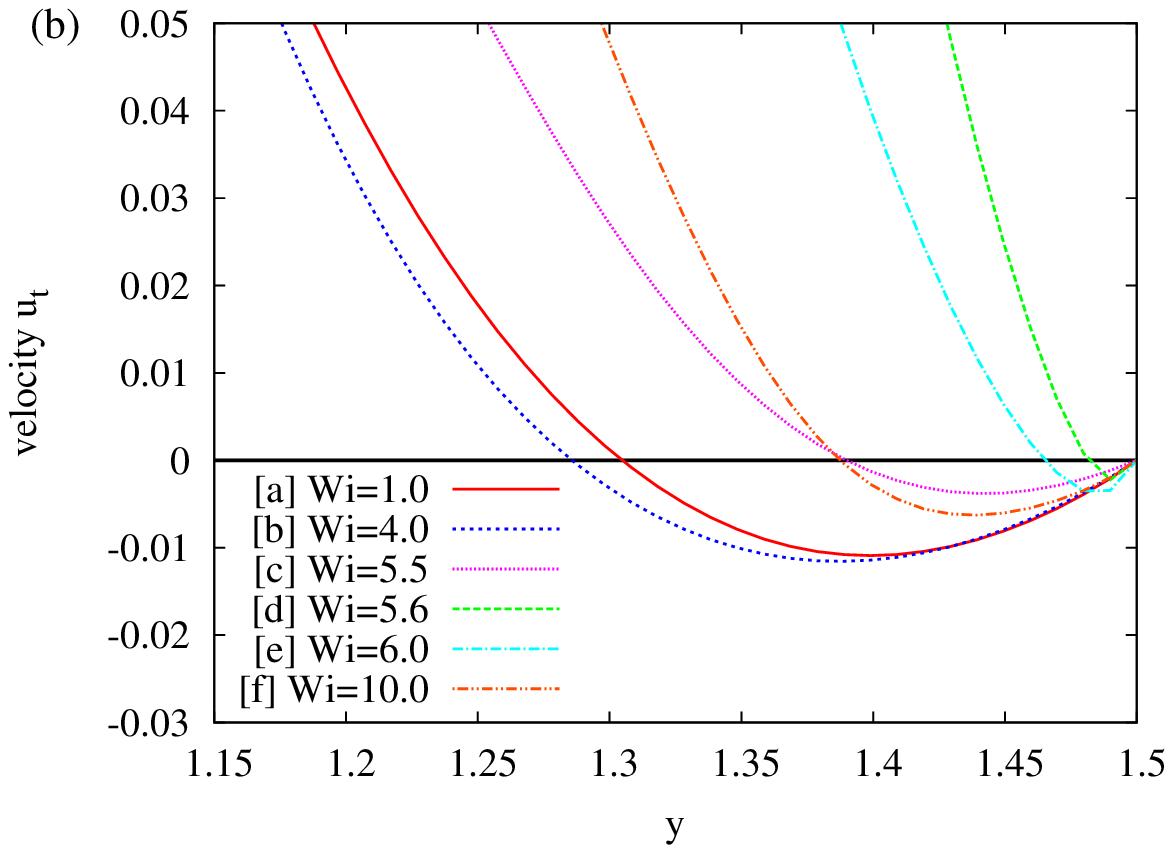}  }
 \caption{(Color online) (a) Tangential velocities $u_t$ in the vertical section at $x=\pi/4$
  in the convective region [a] $\mathrm{Wi}=1.0$ [b] $\mathrm{Wi}=4.0$, [c] $\mathrm{Wi}=5.5$,
  in the transition region [d] $\mathrm{Wi}=5.6$, [f] $\mathrm{Wi}=6.0$, and in the elastic region [e] $\mathrm{Wi}=10.0$ at $\mathrm{Re}=400$. 
  (b) Tangential velocities $u_t$ in the vertical section at $x=\pi/4$ magnified in $1.15\leq y\leq1.5$. }
  \label{fig:flowsect400}
\end{figure}

\begin{figure}
  \centerline{\includegraphics[height=170pt]{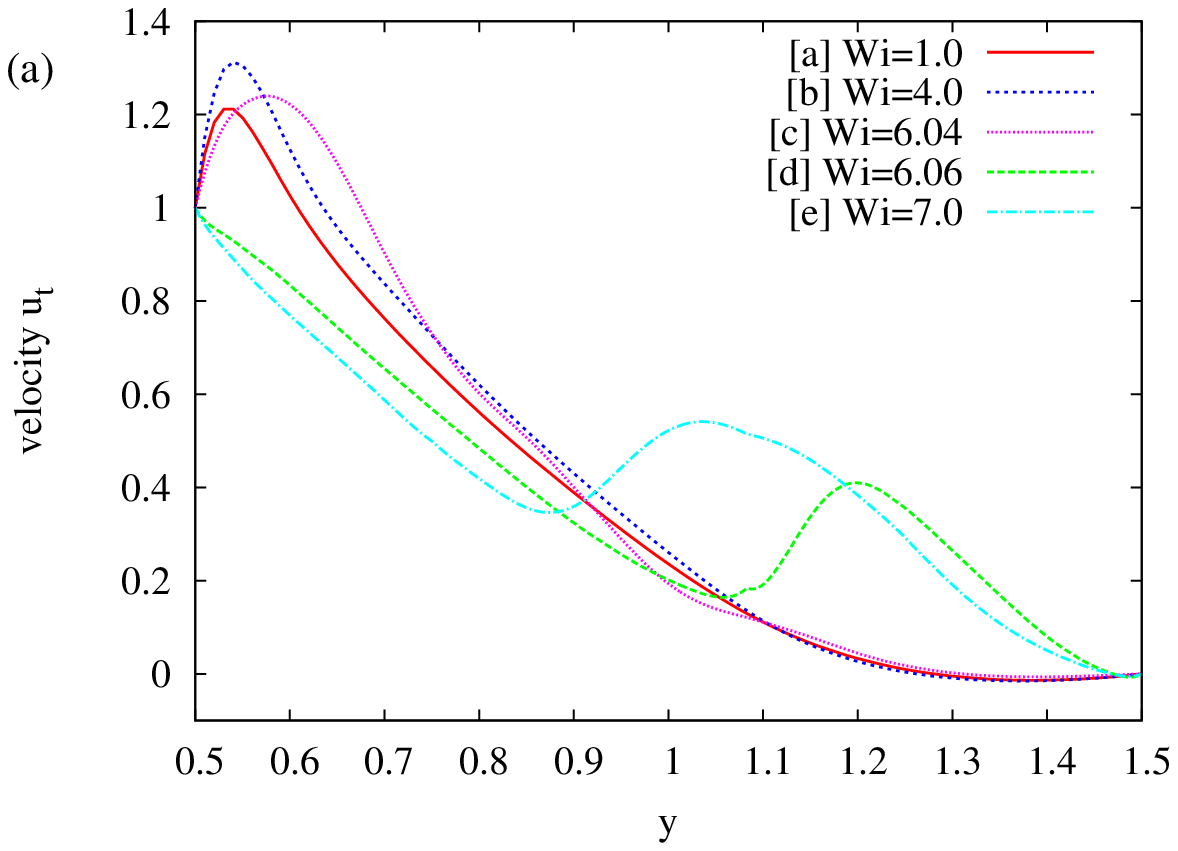}
  \includegraphics[height=170pt]{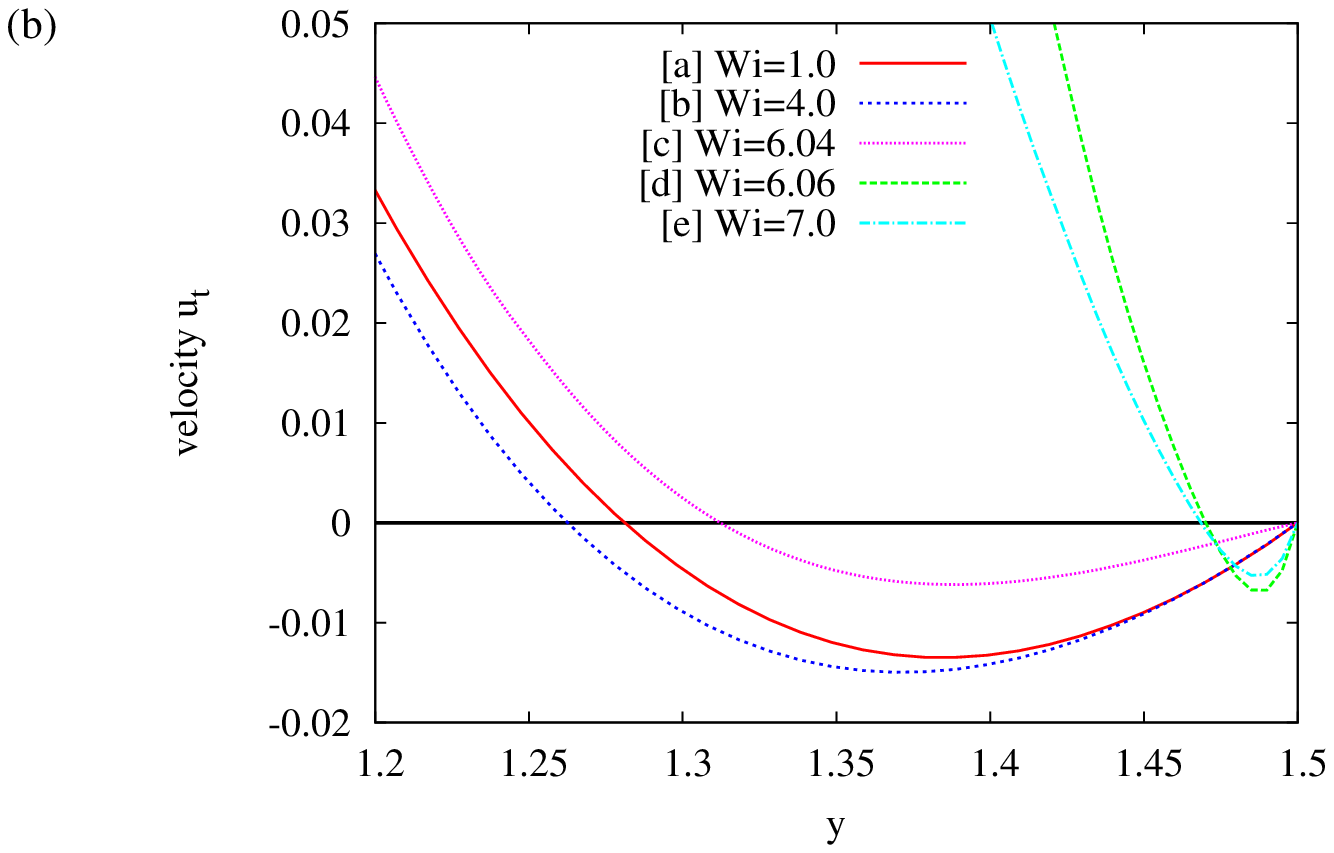} }
  \caption{(Color online) (a) Tangential velocities $u_t$ in the vertical section at $x=\pi/4$
  in the convective region [a] $\mathrm{Wi}=1.0$ [b] $\mathrm{Wi}=4.0$, [c] $\mathrm{Wi}=6.04$,
  in the transition region [d] $\mathrm{Wi}=6.06$, and [e] $\mathrm{Wi}=7.0$ at $\mathrm{Re}=800$. 
  (b) Tangential velocities $u_t$ in the vertical section at $x=\pi/4$ magnified in $1.2\leq y\leq1.5$. }
  \label{fig:flowsect800}
\end{figure}

The other difference of flow patterns in the three regions is "jet flow" observed in the vertical section of flow velocity.
Figures\ \ref{fig:flowsect}, \ref{fig:flowsect400}, and \ref{fig:flowsect800} show the vertical sections of the velocity component $u_x$
at $x=\pi/4$ with $\mathrm{Re}=200$, $400$, and $800$.
We define jet flows the region in which the velocity is larger than in the surrounding area.
Jet flows whose velocity is larger than wall speed (unity) are observed near the bottom wall $0.5<y<0.7$
in the convective region as shown in the curves [a]-[c] in Figs.\ \ref{fig:flowsect}(a), \ref{fig:flowsect400}(a), and \ref{fig:flowsect800}(a).
The jet flows near the bottom wall suddenly disappear at $\mathrm{Wi}_{\mathrm{c}1}$
as shown in the curve [d] in Figs.\ \ref{fig:flowsect}(a), \ref{fig:flowsect400}(a), and \ref{fig:flowsect800}(a).
Simultaneously, jet flows appear near the upper wall as shown in the same velocity profiles.
The jet flows near the upper wall are broader than those near the bottom wall in the convective region.
The jet flows gradually reach to the bottom wall with $\mathrm{Wi}$ during the transition region as shown in the curves [d] and [e]
in Figs.\ \ref{fig:flowsect}(a), \ref{fig:flowsect400}(a), and \ref{fig:flowsect800}(a).
In the elastic region, the jet flows are repeatedly present near the bottom wall
though the jet flows are slower and broader than those in the convective region as shown in the curve [f]
in Fig.\ \ref{fig:flowsect}(a), \ref{fig:flowsect400}(a), and \ref{fig:flowsect800}(a).
The emergence of the jet flows does not depend on $\mathrm{Re}$ except the jet flows at larger $\mathrm{Re}$ is broader in the elastic region.
As described below in detail, the jet flows are also caused by an streamwise elastic force due to polymer stretching in viscoelastic fluid flows.
Note that negative-velocity regions appear near the upper wall as shown in the velocity profiles magnified near the upper wall
in Figs.\ \ref{fig:flowsect}(b), \ref{fig:flowsect400}(b), and \ref{fig:flowsect800}(b).
These negative-velocity regions exist in the separation vortices in the outward region (a) as shown in Figs.\ \ref{fig:streamline200} and \ref{fig:streamline400}.

The jet flows are also observed in the flow-speed profiles as shown in the right figures in Figs.\ \ref{fig:streamline200} and \ref{fig:streamline400}.
In the convective region, the bright regions in the inward region (b) correspond to the jet flows.
In the transition region, the jet flows correspond to the regions where the flow speed is larger than the surrounding area,
from the outward region (a) to the contraction region (c) near the upper wall.
In the elastic region, the jet flows emerge from the contraction region (c)
to the contraction region (f) via the inward region (b) near the bottom wall.

We call the change of flow patterns including the separation vortex and the jet flow in the three regions "flow pattern transition".

\subsection{Characteristics of flows}
\label{sec:characteristics}

In the previous subsection, we described that the steady solutions are classified into the three groups.
The solutions belonging to different groups have remarkably different flow patterns.
How the change of flow patterns influences characteristic quantities of flows such as a wall friction and a flow rate?
In this subsection, we show the dependence of a viscous wall friction, an elastic wall friction, and a flow rate on
$\mathrm{Wi}$ at $\mathrm{Re}=200$ and $400$.

\subsubsection{Laminar drag reduction}
\label{sec:laminardrag}

The flow pattern transition described in the previous subsection
accompanies the abrupt change of a viscous wall friction.
The viscous wall friction is very important for engineering applications to the control of the loss of flow rate.

We define the average viscous wall friction (viscous drag) acting on the walls of the channel as follows:
\begin{equation}
\tau_{vwf} = \frac{1}{S_{\Gamma}}\int_{\Gamma}\beta\left|\frac{\partial u_t}{\partial n}\right|d\Gamma,
\end{equation}
where $\Gamma$ represents the total of the bottom boundary $\Gamma_1$ and the upper boundary $\Gamma_2$.
$S_{\Gamma}$ is the sum of the lengths of the upper and bottom boundary walls.
$t$ and $n$ denote a tangential and a normal directions as shown in Fig.\ \ref{fig:geometry}(a).
$\partial u_t/\partial n$ represent the derivative of the tangential velocity component $u_t$ with respect to the wall-normal direction $n$.

\begin{figure}
  \centerline{\includegraphics[height=200pt]{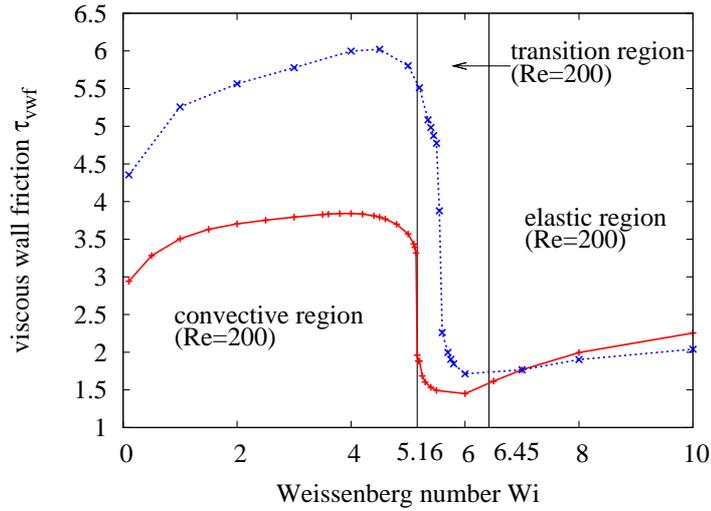}}
  \caption{(Color online) Average viscous wall friction $\tau_{vwf}$ as a function of $\mathrm{Wi}$ at $\mathrm{Re}=200$ and $400$.
  Two vertical lines represent the first and second critical Weissenberg numbers at $\mathrm{Re}=200$.  }
  \label{fig:sfrict}
\end{figure}

Figure \ref{fig:sfrict} shows the average viscous wall friction as a function of $\mathrm{Wi}$ at $\mathrm{Re} = 200$ and $400$.
The viscous drag is abruptly reduced at $\mathrm{Wi}_{\mathrm{c}1}$,
while the viscous drag gradually grows in the convective and elastic regions.
The abrupt reduction of the viscous drag simultaneously occurs with the vanishment of the separation vortex.

We call the drag reduction in laminar steady flows "laminar drag reduction".
The expression has been also used in the studies of viscoelastic fluid flows in curved pipes with circular cross sections \cite{Fan:2001}.

\subsubsection{Flow rate}
\label{sec:flowrate}

The flow rate $FR$ in the channel is defined in the following way:
\begin{equation}
FR = \int_{0}^{1}v_x(\pi/2,y)dy.
\end{equation}

\begin{figure}
  \centerline{\includegraphics[height=200pt]{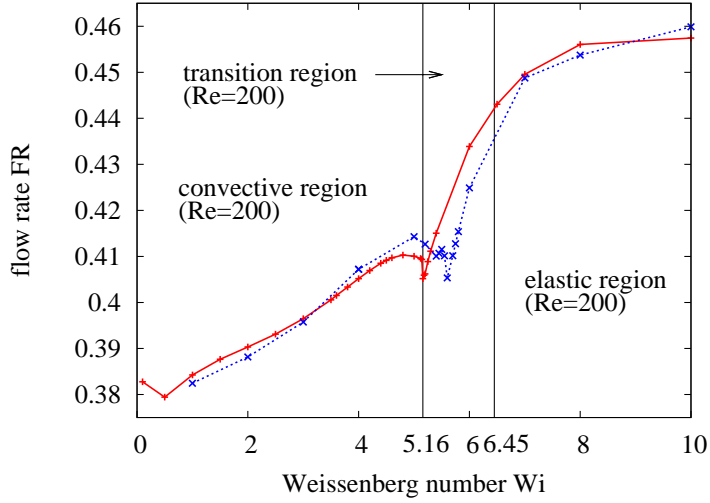}}
  \caption{(Color online) Flow rate $\mathrm{FR}$ as a function of $\mathrm{Wi}$ at $\mathrm{Re} = 200$ and $400$.
  Two vertical lines represent the first and second critical Weissenberg numbers at $\mathrm{Re}=200$.  }
  \label{fig:frate}
\end{figure}

Figure\ \ref{fig:frate} shows the flow rate $FR$ as a function of $\mathrm{Wi}$ at $\mathrm{Re} = 200$ and $400$.
In the convective region, the flow rate gradually increases, while the flow rate is temporarily decreased at $\mathrm{Wi}_{\mathrm{c}1}$.
The flow rate greatly grows in the transition region.
In the elastic region, the growth of the flow rate is relaxed.

\subsubsection{Growth of flow resistance}
\label{sec:resistance}

In this subsection, we evaluate the following average elastic wall friction:
\begin{equation}
\tau_{ewf} = \frac{1}{S_{\Gamma}}\int_{\Gamma}\tau_{tn}d\Gamma,
\label{eq:wallfriction}
\end{equation}
where $\tau_{tn}$ is a component of the extra-stress tensor in the coordinate which consists of two base vectors.
One is a tangential unit vector (denoted as $\bm t$) and the other is a normal unit vector (denoted as $\bm n$) as shown in Fig.\ \ref{fig:geometry}(a).
This elastic wall friction represents a flow resistance resulting from the elastic stress.
The sum of the viscous and elastic wall frictions is a total flow resistance acting on the walls.

\begin{figure}
  \centerline{\includegraphics[height=200pt]{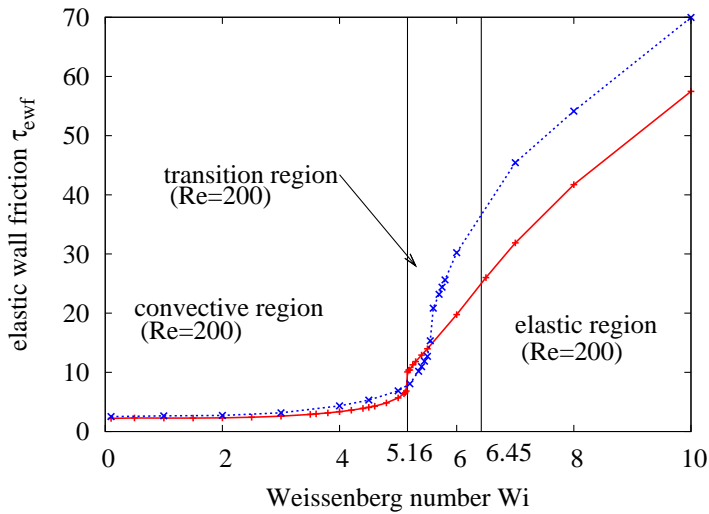}}
  \caption{(Color online) Average elastic wall friction $\tau_{ewf}$ as a function of $\mathrm{Wi}$ at $\mathrm{Re} = 200$ and $400$.
     Two vertical lines represent the first and second critical Weissenberg numbers at $\mathrm{Re}=200$.  }
  \label{fig:frict}
\end{figure}

Figure\ \ref{fig:frict} shows the elastic wall friction as a function of $\mathrm{Wi}$ at $\mathrm{Re} = 200$ and $400$.
At $\mathrm{Wi}_{\mathrm{c}1}$, the elastic wall friction suddenly grows.
The sudden growth of the flow resistance also simultaneously occurs with the vanishment of the separation vortex described at Sec.\ \ref{sec:classification}.

\subsection{Contribution of elastic force onto the flow pattern transition}
\label{sec:balance}

In this subsection, we first introduce the norm of the absolute value of each term in the Navier-Stokes equation.
We next elucidate the mechanism of the flow pattern transition by utilizing the relation among the norms.

The norm of the absolute value of each term is defined in the following way:
\begin{equation}
\|\mathrm{Re}({\bm u}\cdot\nabla){\bm u}\| = \int_{\Omega}|\mathrm{Re}({\bm u}\cdot\nabla){\bm u}|d\Omega.
\end{equation}

\begin{figure}
  \centerline{\includegraphics[height=200pt]{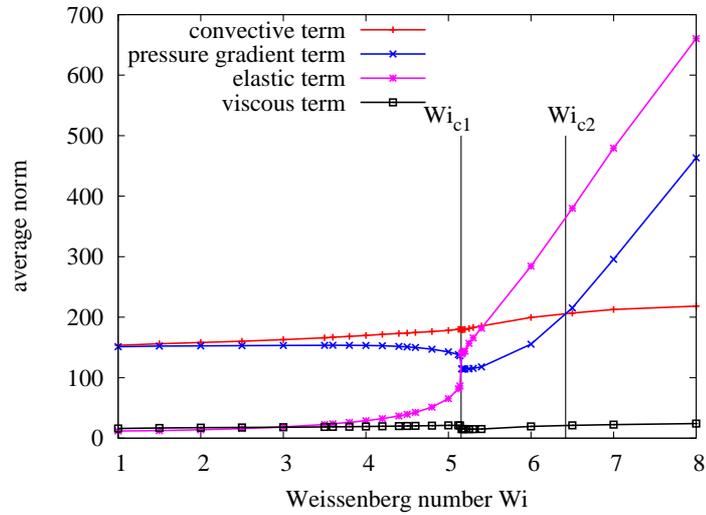}}
  \caption{(Color online) Absolute-value norms of the convective term $\|\mathrm{Re}({\bm u}\cdot\nabla){\bm u}\|$,
    the pressure gradient term $\|\nabla p\|$, the elastic term $\|\nabla\cdot{\bm\tau}\|$,
       and the viscous term $\|\beta\nabla^2{\bm u}\|$ in the Navier-Stokes equation as a function of $\mathrm{Wi}$ at $\mathrm{Re}=200$. }
  \label{fig:balance200}
\end{figure}

Figure\ \ref{fig:balance200} shows the absolute-value norms of the convective term $\|\mathrm{Re}({\bm u}\cdot\nabla){\bm u}\|$,
the pressure gradient term $\|\nabla p\|$, the elastic term $\|\nabla\cdot{\bm\tau}\|$,
and the viscous term $\|\beta\nabla^2{\bm u}\|$ in the Navier-Stokes equation as a function of $\mathrm{Wi}$ at $\mathrm{Re}=200$.

We define the first critical Weissenberg number $\mathrm{Wi}_{\mathrm{c}1}$ as the Weissenberg number
at which the norm of the elastic term intersects with that of the pressure gradient term.
The second critical Weissenberg number $\mathrm{Wi}_{\mathrm{c}2}$ is similarly defined as the Weissenberg number
at which the norm of the pressure gradient term crosses with that of the convective term.

\begin{figure}
  \centerline{\includegraphics[height=200pt]{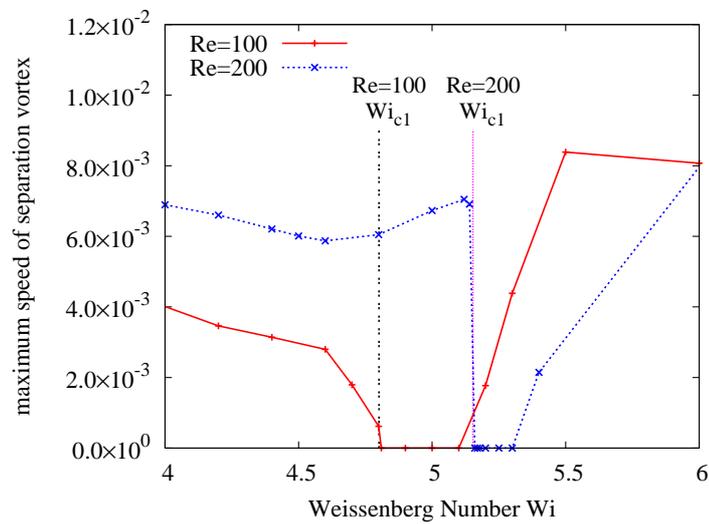}}
  \caption{(Color online) Maximum speed of flows in the separation vortex as a function of $\mathrm{Wi}$ at $\mathrm{Re}=100$ and $200$.
    The vertical lines denote the first critical Weissenberg number $\mathrm{Wi}_{\mathrm{c}1}$ at $\mathrm{Re}=100$ and $200$.}
  \label{fig:vorvelo}
\end{figure}

We calculate the maximum speed of flows in the separation vortex as an intensity of the separation vortex.
The maximum speed is defined in the following way.
We consider the region where $u_x<0$ in the channel.
This region exists in the separation vortex as shown in Figs.\ \ref{fig:streamline200}(a) and \ref{fig:flowsect}(b).
We calculate the maximum value of the flow speed $|{\bm u}|$ in this region.
When the separation vortex is vanished, the region where $u_x<0$ also disappears.
Then, the maximum speed is set to zero.
In Fig.\ \ref{fig:vorvelo}, we show the maximum speed as a function of $\mathrm{Wi}$ at $\mathrm{Re}=100$ and $200$.
This figure shows that the first critical Weissenberg number $\mathrm{Wi}_{\mathrm{c}1}$ coincides with the Weissenberg number
at which the separation is suddenly depressed.

The second critical Weissenberg number provides an estimate of the Weissenberg number at which the jet flows reach to the bottom wall in the transition region,
though the estimate is less exact than the accordance of $\mathrm{Wi}_{\mathrm{c}1}$ to the Weissenberg number at which the separation vortex is depressed.

\begin{figure}
  \centerline{\includegraphics[height=400pt]{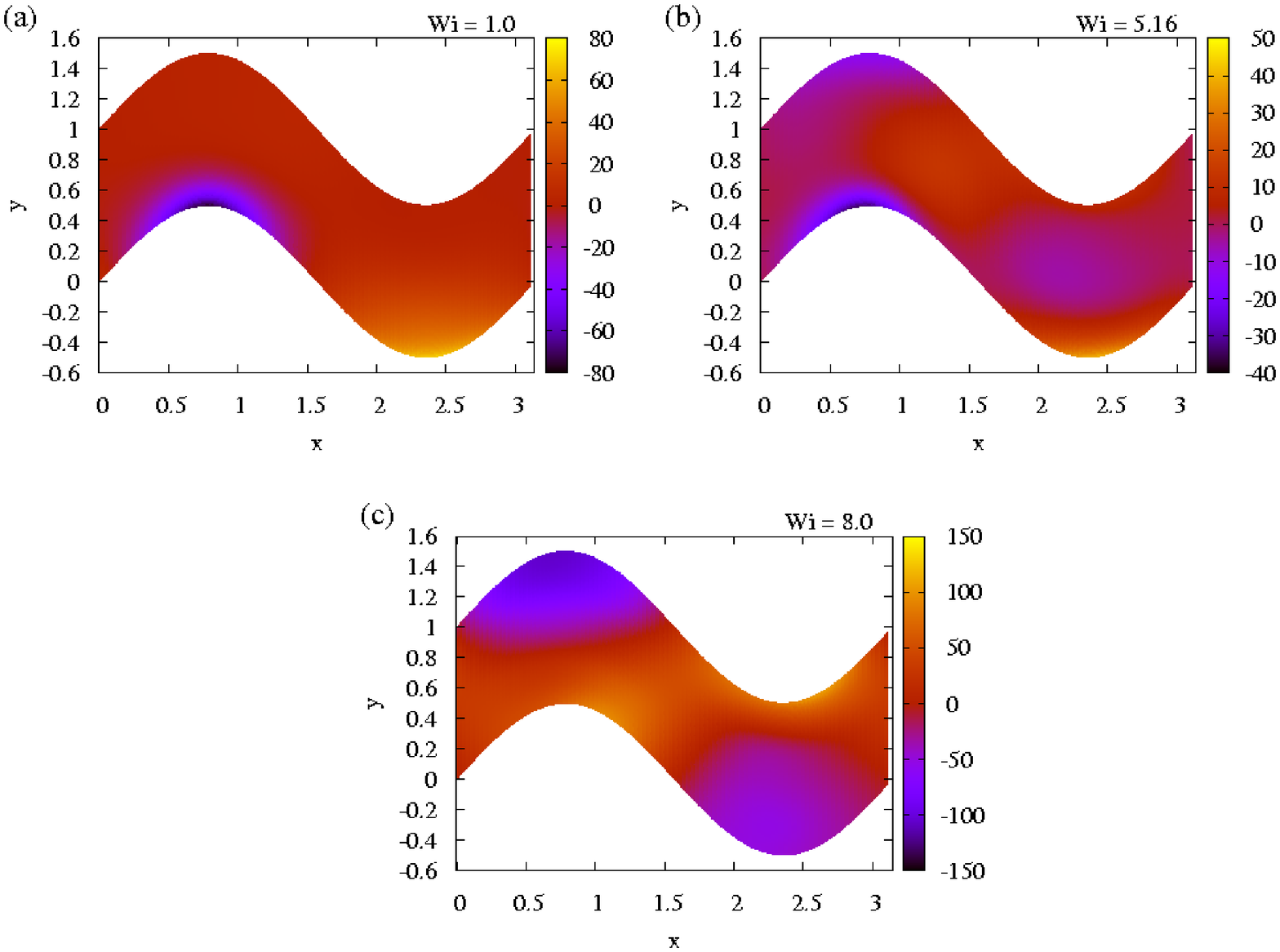} }
  \caption{(Color online) Pressure profiles in (a) the convective region $\mathrm{Wi}=1.0$,
   (b) the transition region $\mathrm{Wi}=5.16$ and (c) the elastic region $\mathrm{Wi}=8.0$ at $\mathrm{Re}=200$.}
  \label{fig:pressure200}
\end{figure}

\begin{figure}
  \centerline{\includegraphics[height=550pt]{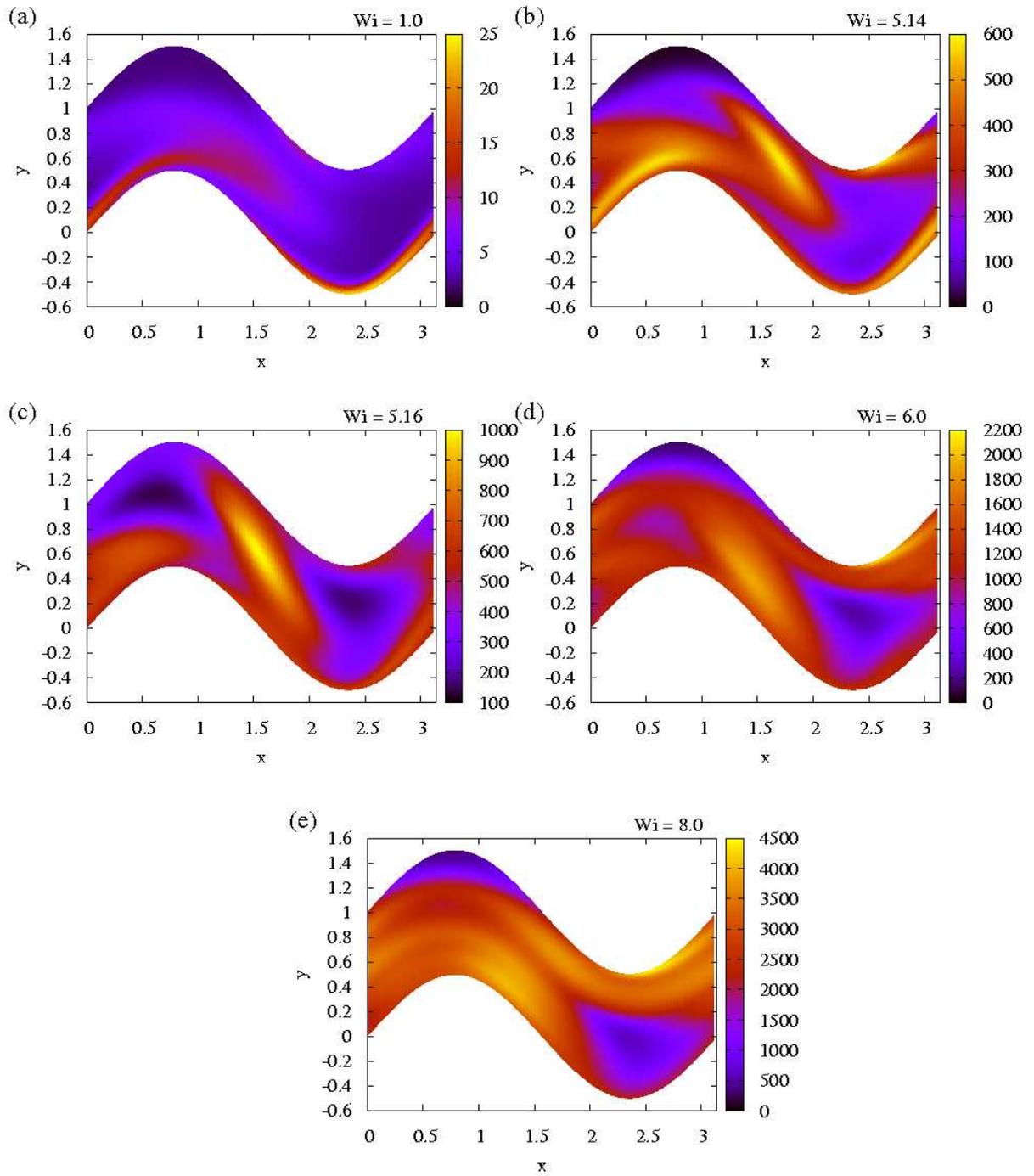} }
  \caption{(Color online) Profiles of the trace of the conformation tensor $\mathrm{Tr}({\bm C})$ in the convective regions (a) $\mathrm{Wi}=1.0$, (b) $\mathrm{Wi}=5.14$,
      in the transition regions (c) $\mathrm{Wi}=5.16$ , (d) $\mathrm{Wi}=6.0$ and in the elastic region (e) $\mathrm{Wi}=8.0$ at $\mathrm{Re}=200$.}
  \label{fig:conform200}
\end{figure}

\begin{figure}
  \centerline{\includegraphics[height=350pt]{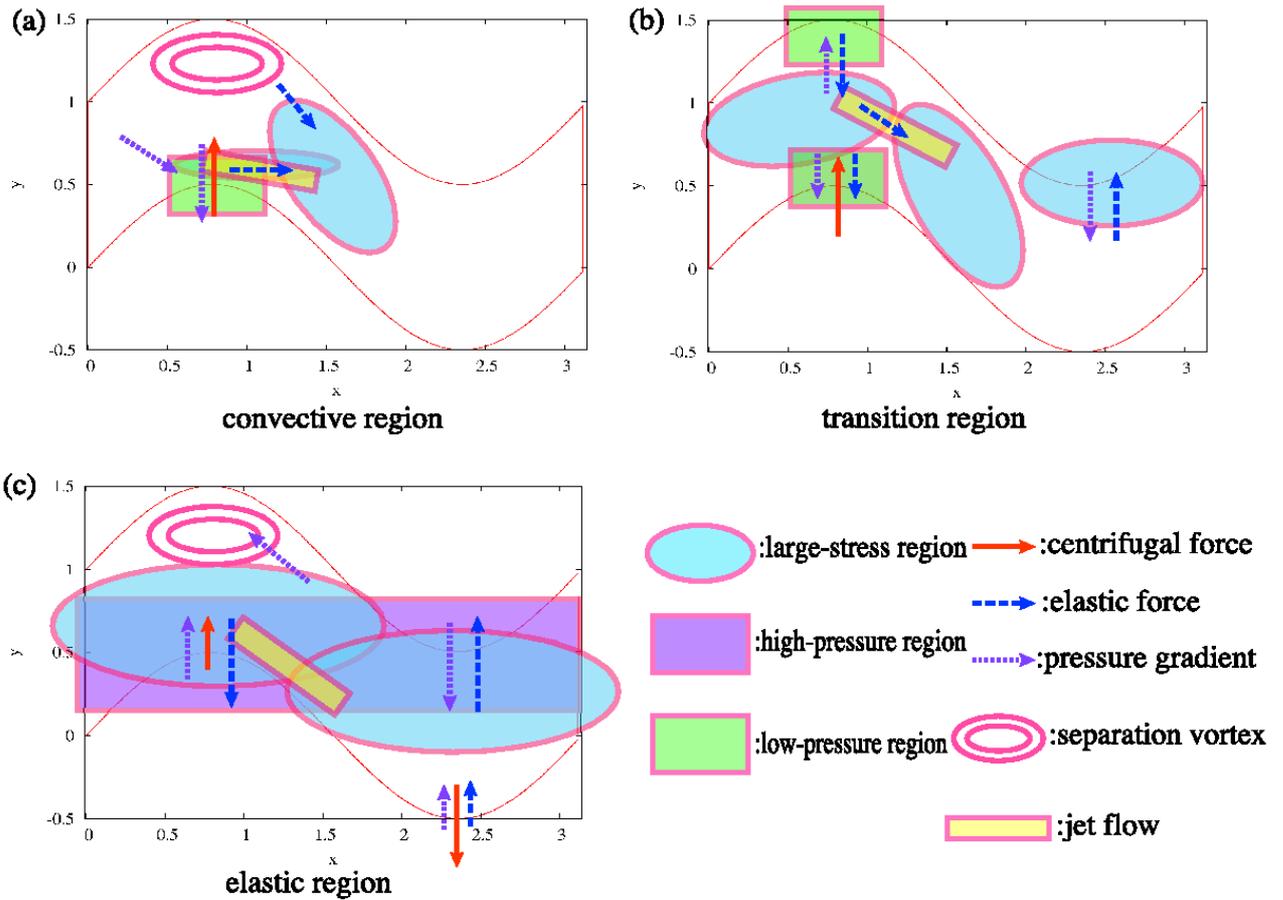} }
  \caption{(Color online) Diagrams schematically representing the local relation among the centrifugal force, the elastic force,
  and the pressure gradient in the (a) convective, (b) transition, and (c) elastic regions.
  The characteristic regions and patterns used in the description at Sec.\ \ref{sec:balance} are superposed. }
  \label{fig:forceschem}
\end{figure}

We subsequently describe the mechanism of the flow pattern transition.
Here, we utilize the profiles of the pressure and the conformation tensor as well as the norms of the terms for the description.
Figures\ \ref{fig:pressure200} and \ref{fig:conform200} display the profiles of the pressure and the trace of the conformation tensor $\mathrm{Tr}({\bm C})$
in the convective, transition, and elastic regions at $\mathrm{Re}=200$.
The trace of the conformation tensor represents the magnitude of polymer stretching.
Note that the pressure is fixed to zero at the given point.
Negative pressure at certain point then represents the pressure smaller than at the reference point.
In these figures, we can see the local relation among elastic force, pressure gradient, and centrifugal force
defined as $\mathrm{Re}(u_t^2/R)$ where $R$ is the curvature radius of streamlines.
Here, we notice that the streamwise elastic force plays a key role in the flow pattern transition as well as the hoop stress.
Figure\ \ref{fig:forceschem} schematically summarizes the transition of the local relation among the forces in the three regions.
At the subsequent subsections, we describe the flow pattern transition in the three regions by utilizing these quantities.

\subsubsection{Convective region}
\label{sec:convective}

In the convective region, the separation vortex emerges in the outward region (a) (see Fig.\ \ref{fig:streamline200}(a)).
This is a characteristic phenomenon of Newtonian fluid flows in wavy-walled channels \cite{Asako:1988}. 

The generation mechanism of the separation vortex is as follows.
The norms of the convective and pressure-gradient terms almost accord in the convective region as shown in Fig.\ \ref{fig:balance200},
while the elastic and viscous terms are much less than the convective and pressure-gradient terms.
This accordance indicates that the pressure gradient and the centrifugal force cancel in the whole domain in steady flows.
The centrifugal force is very large particularly in the inward region (b) where the streamwise velocity and the curvature are large.
The large centrifugal force leads to the large centripetal pressure gradient in the inward region (b).
Then, the pressure becomes smaller in the inward region (b) than in the outward region (a) (see Fig.\ \ref{fig:pressure200}(a)).
Because the pressure gradient inhibits the flows from the contraction region (f) to the outward region (a),
the bulk flow then passes near the inward region (b).
The separation vortex detached from the bulk flow finally emerges in the outward region (a) (see Fig.\ \ref{fig:streamline200}(a)).

In the convective region, the characteristic phenomenon of viscoelastic fluid flows is observed as a jet flow near the bottom wall
as shown in Figs.\ \ref{fig:flowsect}(a), \ref{fig:flowsect400}(a), and \ref{fig:flowsect800}(a).
The generation mechanism of the jet flow is described as follows.
The bulk shear flow through the inward region (b) strongly stretches polymers near the bottom wall.
The polymer stretching leads to the increase of the elastic stress and to that of the streamwise elastic force in this region (see Fig.\ \ref{fig:conform200}(a)).
Because the augmented streamwise elastic force increases a streamwise flow,
the jet flow finally emerges in the inward region (b) (see $0.5<y<0.7$ in the curves [a]-[c] in Fig.\ \ref{fig:flowsect}(a)).

The jet flow causes the viscous drag enhancement in the convective region as described at Sec.\ \ref{sec:laminardrag} in the following way.
Due to the jet flow, the wall-normal velocity gradient is inverted near the bottom wall.
The elastic force near the bottom wall gradually increases with $\mathrm{Wi}$ (see Fig.\ \ref{fig:frict}).
Then, the velocity of the jet flow is also augmented due to the intensified elastic force as shown in the curves [a] and [b] in Fig.\ \ref{fig:flowsect}(a).
Because the increase of the velocity leads to that of the velocity gradient,
the viscous wall friction also grows on the bottom wall as shown in Fig.\ \ref{fig:sfrict}.
The increase of the flow velocity due to the streamwise elastic force near the bottom wall
also directly leads to the growth of the flow rate as shown in Fig.\ \ref{fig:frate}.

With further increasing $\mathrm{Wi}$, the flow behavior described above is moderately modified.
The representative modification is the growth of the elastic stress in the contraction regions (c) and (f) (see Fig.\ \ref{fig:conform200}(b)).
In the contraction regions (c) and (f), the channel is narrower than in the highly-curved regions
including the outward regions (a), (e), the inward regions (b), and (d) (see Fig.\ \ref{fig:geometry}(a)).
The flow velocity is basically large in the narrow regions (c) and (f)
because the flow rate, which is the integral of the flow velocity in the wall-normal section of the channel, is constant in the every wall-normal section.
The difference between the velocities in the contraction region and in the highly-curved region stretches polymers between the two regions.
The augmented elastic stress forms the lumps of large-stress regions in the contraction regions (c) and (f) (see Fig.\ \ref{fig:conform200}(b)).
The lump in the contraction region (c) gradually extends to the outward regions (a) and (e) with $\mathrm{Wi}$
and the lump in the contraction region (f) gradually reaches to the inward region (b) (see Fig.\ \ref{fig:conform200}(b)).
Because the lump of the elastic stress induces flows from the outward region (a) to the contraction region (c)
due to the tangential gradient of the elastic stress (see the definition of the tangential direction in Fig.\ \ref{fig:geometry}(a)),
the separation vortex in the outward region (a) gradually becomes small with $\mathrm{Wi}$ (see the curves [b] and [c] in Fig.\ \ref{fig:flowsect}(b)).
Since the large-stress region extended from the contraction region (f) to the inward region (b) increases flows above the jet flow in the inward region (b),
the jet flow is relatively depressed (see $0.5<y<0.7$ in the curves [b] and [c] in Fig.\ \ref{fig:flowsect}(a)).
The depression of the jet flow directly leads to the decrease of the viscous wall friction on the bottom wall at $\mathrm{Wi}>4.0$ as shown in Fig.\ \ref{fig:sfrict}.

The description at this subsection is summarized as follows (see also Fig.\ \ref{fig:forceschem}(a)).
The separation vortex in the outward region (a) and the jet flow in the inward region (b)
are the characteristic flow patterns of the steady solutions in the convective region.
The separation vortex is caused by the local forces which also works in the Newtonian fluid flows,
whereas the jet flow stems from the elastic force characteristic of viscoelastic fluid flows.

\subsubsection{Transition region}
\label{sec:transition}

As described at the previous subsection, the flow from the outward region (a) to the contraction region (c)
gradually increases with $\mathrm{Wi}$ due to an streamwise elastic force.
The increased flow further augments the elastic stress in the contraction region (c)
because the increased velocity and velocity gradient further stretch polymers.
The augmented elastic stress repeatedly increases the flow from the outward region (a) to the contraction region (c).
This argument indicates that the flow and the streamwise elastic force is self-enhanced in this region.

When $\mathrm{Wi}$ attains the first critical Weissenberg number $\mathrm{Wi}_{\mathrm{c}1}$,
this self-enhancing procedure actually sets in.
Then, the elastic effect becomes suddenly comparable with the convective effect,
and the flow pattern observed in the convective region drastically changes.

The sudden formation of the jet flow in the outward region (a) and the sudden vanishment of the jet flow in the inward region (b)
are the prominent change of the flow pattern in the transition region as described at Sec.\ \ref{sec:classification}
(see $0.5<y<0.7$ and $1.1<y<1.5$ in the curves [c] and [d] in Fig.\ \ref{fig:flowsect}(a)).
The generation of the jet flow in the outward region (a) results from the increase of the velocity in the outward region (a)
due to the streamwise elastic force described above.
Then, the jet flow in the inward region (b) is relatively depressed due to the increase of flows to the outward region (a).
The vanishment of the jet flow in the inward region (b) leads to the abrupt reduction of the viscous drag on the bottom wall as shown in Fig.\ \ref{fig:sfrict}. 
Due to the emergence of the jet flow in the outward region (a), the separation vortex is vanished or becomes very small
as shown in Fig.\ \ref{fig:streamline200}(b) and $y > 1.4$ in the curve [d] in Fig.\ \ref{fig:flowsect}(a).

The lump of the large-stress region in the contraction region (c) also suddenly grows beyond $\mathrm{Wi}_{\mathrm{c}1}$.
Because the lump of the large-stress region attains to the walls (see Fig.\ \ref{fig:conform200}(c)),
the elastic wall friction suddenly grows at $\mathrm{Wi}_{\mathrm{c}1}$ as shown in Fig.\ \ref{fig:frict}.
Note that the lump of the elastic stress is inclined to the channel walls.
This is because the jet flow in the outward region (a) enhances the elastic stress near the upper wall (see Figs.\ \ref{fig:conform200}(c)).

The flow rate for $\mathrm{Wi}>4.0$ varies with the decrease of the flow velocity in the inward region (b)
and the increase of the flow velocity in the outward region (a).
The flow pattern transition leads to the reduction of the flow rate as shown in Fig.\ \ref{fig:frate}.
However, the flow rate remarkably increases above $\mathrm{Wi}_{\mathrm{c}1}$
because the streamwise elastic force further increases the flow to the outward region (a) with $\mathrm{Wi}$ in the transition region.
The growth of the flow rate with $\mathrm{Wi}$ also appears in the amplitude of the jet flow as shown in the curves [d]-[f] in Fig.\ \ref{fig:flowsect}(a).

In the transition region, the jet flow in the outward region (a)
gradually moves from the upper wall to the bottom wall (see the curves [d]-[f] in Fig.\ \ref{fig:flowsect}(a)).
The mechanism of the moving jet flow is described as follows.
The hoop stress induces the elastic force working to the centripetal direction in curved streamlines.
Due to the centripetal elastic force in the outward region (a), flows normal to the streamwise direction are induced.
Because of the flow normal to the streamwise direction, the lump of the large-stress region in the contraction region (c)
falls down to the bottom wall (see Fig.\ \ref{fig:conform200}(d)).
The descended lump of the large-stress region repeatedly induces the jet flow due to the streamwise elastic force in the lower position of the channel
(see Fig.\ \ref{fig:conform200}(d) and the curve [e] in Fig.\ \ref{fig:flowsect}(a)).
Because the jet flow augments the elastic stress between the outward region (b) and the inward region (a),
the large-stress belt-like region is finally formed from the region between the outward region (b) and the inward region (a)
to the inward region (d) (see Fig.\ \ref{fig:conform200}(d)).
The hoop stress strongly works in this region due to the large elastic stress and the large flow curvature.
Then, the flow normal to the streamwise direction is further induced.
Thus, the jet flow gradually reaches to the bottom wall.
With the variation of the position of the jet flow, the separation vortex repeatedly emerges and gradually grows
(see the curves [d]-[f] in Fig.\ \ref{fig:flowsect}(b)).

The influence of the elastic stress on the flow is confirmed through the comparison of the elastic force
with the centrifugal force and the pressure gradient in the following way (see also Fig.\ \ref{fig:forceschem}(b)).
First, the hoop stress in the outward region (a) described above is canceled by
the centrifugal pressure gradient in the steady flows because the centrifugal force is very small due to the small velocity near the upper wall.
The pressure distribution shown in Fig.\ \ref{fig:pressure200}(b) indeed indicates the centrifugal pressure gradient in the outward region (a).
Second, in the inward region (b), the sum of the centripetal elastic force and pressure gradient cancels the centrifugal force in the steady flows
because the convective term, the origin of the centrifugal force, is yet larger than the elastic and pressure gradient terms
at relatively small $\mathrm{Wi}$ in the transition region as shown in Fig.\ \ref{fig:balance200}.
Since the centrifugal force does not almost depends on $\mathrm{Wi}$ (see Fig.\ \ref{fig:balance200}),
the growth of the elastic force directly leads to the decrease of the pressure gradient in the inward region (b)
(see Figs.\ \ref{fig:pressure200}(a) and (b)).
The reduction of the pressure gradient in the inward region (b) is also consistent with the decrease of the flow to the inward region (b).

The description at this subsection is summarized as follows (see also Fig.\ \ref{fig:forceschem}(b)).
The sudden vanishment of the jet flow in the inward region (b)
and the abrupt formation of the jet flow in the outward region (a) due to the streamwise elastic force simultaneously occurs at the beginning of the transition region.
The jet flow gradually reaches from the upper wall to the bottom wall due to the elastic hoop stress in the transition region.

\subsubsection{Elastic region}
\label{sec:elastic}

The jet flow approaching to the bottom wall stops approximately above the second critical Weissenberg number.
In the elastic region, the jet flow near the bottom wall and the separation vortex in the outward region (a)
characterizes the flow patterns as in the convective region (see Figs.\ \ref{fig:streamline200}(a) and (c)),

Despite the similarity of the flow patterns between the convective and elastic regions,
the generation mechanism of the separation vortex and the jet flow is remarkably different in the two regions.
In the elastic region, the large-stress belt-like region described at the previous subsection covers wide regions including the inward regions (b), (d)
, the contraction regions (c), and (f) in the channel due to the large elastic hoop stress at large $\mathrm{Wi}$ (see Fig.\ \ref{fig:conform200}(e)).
Because the convective term does not vary with $\mathrm{Wi}$ as described above,
the pressure gradient term grows with the elastic stress in the steady flows in order to cancel the elastic term.
The large-pressure region thus also spreads into the wide regions of the domain (see Fig.\ \ref{fig:pressure200}(c)).
Then, the flow from the outward region (a) to the contraction region (c) is severely inhibited by the large pressure gradient
from the contraction region (c) to the outward region (a).
Since the bulk flow cannot pass through the outward region (a), the separation vortex is stably maintained in the outward region (a).
Note that the separation vortex appears in the low-pressure region in the elastic region
unlike the separation vortex observed in the high-pressure region in the convective region
(see Figs.\ \ref{fig:streamline200}(a),(c), Figs.\ \ref{fig:pressure200}(a), and (c)).
The vortex emergence in the low-pressure region is a characteristic phenomenon of viscoelastic fluid flows.

The difference of flows between the convective and elastic regions
also arises in the viscous wall friction and the flow rate (see Figs.\ \ref{fig:sfrict} and \ref{fig:frate}).
In the elastic region, the flow is basically developed in the inward regions due to the large elastic hoop stress.
The larger streamwise elastic force at larger $\mathrm{Wi}$ causes the larger velocity of the jet flow and the larger velocity gradient
in the inward region (b) (see Fig. \ref{fig:flowsect}(c)). 
The variation of the velocity gradient directly leads to the viscous drag enhancement with $\mathrm{Wi}$ in the elastic region as shown in Fig.\ \ref{fig:sfrict}.
Because the velocity gradient in the inward region (b) in the elastic region is much smaller than that in the convective region,
the magnitude of the viscous drag in the elastic region is also much smaller than that in the convective region
(see the curves [a]-[c] and [f] in Fig.\ \ref{fig:flowsect}(a)).
The flow rates in the convective and elastic regions are also different
because the magnitudes of the elastic stress in the two regions are remarkably different (see Figs.\ \ref{fig:conform200}(a) and (e)).
However, the growth rate of the flow rate is relaxed with $\mathrm{Wi}$ in the elastic region.
This is because the elastic stress is also relaxed for sufficiently large $\mathrm{Wi}$ in the elastic region as shown in Fig.\ \ref{fig:frict}.

To summarize, the separation vortex in the low-pressure region and the jet flow in the inward regions (b)
emerge due to the large elastic hoop stress and the streamwise elastic force in the elastic region (see also Fig.\ \ref{fig:forceschem}(c)).

We have described the mechanism of the flow pattern transition, the variation of the wall frictions, and that of the flow rate in the three regions.
These arguments show that the streamwise elastic force produces jet flows which characterize the flow pattern formation in wavy-walled channel shear flows.

\subsection{Scaling analysis on the elastic wall friction}
\label{sec:scaling}

How is the relationship between the flow pattern transition and the change of the local force balance theoretically confirmed?
We perform several scaling analyses on the basis of the results described above for theoretical evaluation
because it is severely difficult to derive analytical solutions in our flow system.

\begin{figure}
  \centerline{\includegraphics[height=200pt]{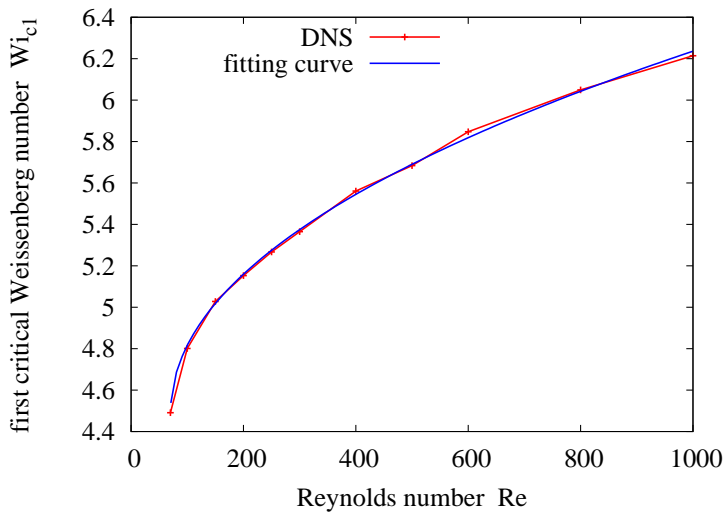}}
  \caption{(Color online) First critical Weissenberg number $\mathrm{Wi}_{\mathrm{c}1}$ as a function of $\mathrm{Re}$.
  The curve of $\mathrm{Wi}_{\mathrm{c}1}$ is fitted with a power function $F_1(\mathrm{Re})=A_1(\mathrm{Re}-\mathrm{Re}_{\mathrm{c}})^{A_2}+A_3$
  where $\mathrm{Re}_{\mathrm{c}}=70$, $A_1\sim0.06$, $A_2\sim0.5$ and $A_3\sim4.5$.}
  \label{fig:rewiwidth}
\end{figure}

In Fig.\ \ref{fig:rewiwidth}, we show the first critical Weissenberg number $\mathrm{Wi}_{\mathrm{c}1}$ as a function of $\mathrm{Re}$.
The curve of $\mathrm{Wi}_{\mathrm{c}1}$ is fitted by using a power function $F_1(\mathrm{Re})=A_1(\mathrm{Re}-\mathrm{Re}_c)^{A_2}+A_3$,
where $\mathrm{Re}_c$ is fixed to $70$.
The values of the fitting parameters are $A_1\sim0.06$, $A_2\sim0.5$, and $A_3\sim4.5$.
The relationship $\mathrm{Wi}_{\mathrm{c}1} \propto \sqrt{\mathrm{Re}-\mathrm{Re}_{\mathrm{c}}}$ is nearly satisfied.
$\mathrm{Re}_{\mathrm{c}}=70$ is a reasonable value because the sudden growth of the elastic wall friction does not occur when $\mathrm{Re}<\mathrm{Re}_{\mathrm{c}}$.
Moreover the size of the separation vortex in the convective region is abruptly decreased when $\mathrm{Re}<\mathrm{Re}_{\mathrm{c}}$.
Although we draw a line from $\mathrm{Re}=70$ to $\mathrm{Re}=0$ in a low $\mathrm{Wi}$ and low $\mathrm{Re}$ region in Fig.\ \ref{fig:phase},
the Reynolds number below which the separation vortex completely disappears in the Newtonian limit $\mathrm{Wi}\to0$ is maybe larger than zero.

The norms of the convective term $\|\mathrm{Re}({\bm u}\cdot\nabla){\bm u}\|$
and the elastic term $|\nabla\cdot{\bm\tau}|$ have similar values at $\mathrm{Wi}_{\mathrm{c}1}$ as shown in Fig.\ \ref{fig:balance200}.
When the non-dimensional velocity and the derivatives are scaled to $O(1)$,
the following relationship is derived:
\begin{equation}
 {\bm\tau} \propto \mathrm{Re}.
 \label{eq:linearrel}
\end{equation}

\begin{figure}
  \centerline{\includegraphics[height=200pt]{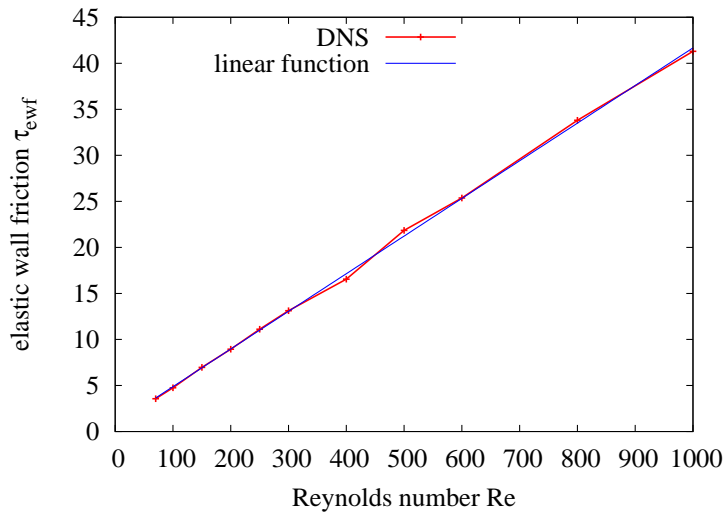}}
  \caption{(Color online) Elastic wall friction $\tau_{ewf}$ at $\mathrm{Wi}_{\mathrm{c}1}$ as a function of $\mathrm{Re}$.
  The linear line $\mathrm{F_2} = B_1(\mathrm{Re}-\mathrm{Re}_c)+B_2$ where $\mathrm{Re}_c=70$, $B_1=4.1\times10^{-2}$, $B_2=3.6$ is superposed.  }
  \label{fig:retauwidth}
\end{figure}

Figure\ \ref{fig:retauwidth} shows the elastic wall friction $\tau_{ewf}$ at $\mathrm{Wi}_{\mathrm{c}1}$ as a function of $\mathrm{Re}$.
We superpose the linear line $\mathrm{F_2} = B_1(\mathrm{Re}-\mathrm{Re}_c)+B_2$
where $\mathrm{Re}_{\mathrm{c}}=70$, $B_1=4.1\times10^{-2}$, $B_2=3.6$ on Fig.\ \ref{fig:retauwidth}.
The linear relationship (\ref{eq:linearrel}) between $\bm\tau$ and $\mathrm{Re}$ is indeed approximately satisfied.
The two relationships ${\bm\tau}\propto\mathrm{Re}-\mathrm{Re}_c$
and $\mathrm{Wi}_{\mathrm{c}1}\propto\mathrm{Re}-\mathrm{Re}_c$ lead to the relationship of the elastic stress and $\mathrm{Wi}_{\mathrm{c}1}$
as ${\bm\tau}\propto{\mathrm{Wi}_{\mathrm{c}1}}^2$.

\subsection{purely elastic limit}
\label{sec:purelyelastic}

In this subsection, we show the results of purely elastic calculations at $\mathrm{Re}=0$.
Recent years, purely elastic flows have been intensively studied in the context of elastic instability and elastic turbulence
\cite{Groisman2:2004,Pakdel:1996,Larson:1990,Thomases:2009}.
The elastic force mainly drives flows in the purely elastic limit because the convective term is vanished.

\begin{figure}
  \centerline{\includegraphics[height=200pt]{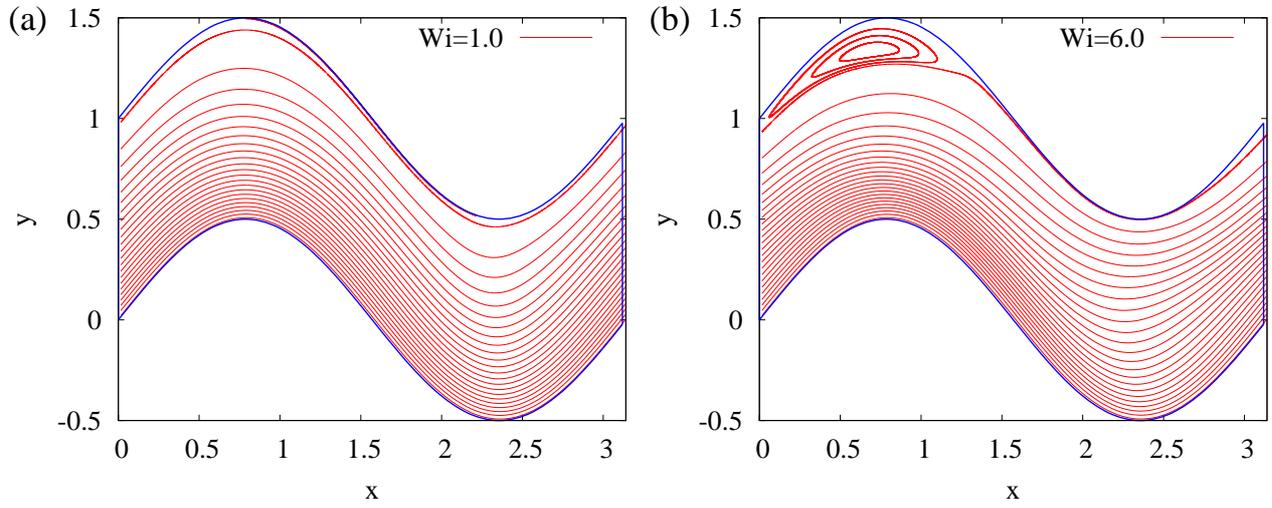} }
  \caption{(Color online) Streamlines of the steady solutions at the small Weissenberg number $\mathrm{Wi}=1.0$
  and the large Weissenberg number $\mathrm{Wi}=6.0$ with $\mathrm{Re}=0$.}
  \label{fig:streamline0}
\end{figure}

Figure\ \ref{fig:streamline0} shows streamlines of the steady solutions
at the low Weissenberg number $\mathrm{Wi}=1.0$ and the moderate Weissenberg number $\mathrm{Wi}=6.0$ with $\mathrm{Re}=0$.
The separation vortex is not observed at low $\mathrm{Wi}$.
The disappearance of the separation vortex results from the vanishment of the convective term.
When the convective term is vanished, the centrifugal force does not work in the whole domain.
Because the centrifugal force is canceled by the pressure gradient for low $\mathrm{Wi}$ as described at Sec.\ \ref{sec:convective},
the pressure gradient is also remarkably weakened in the whole domain.
The depression of the pressure gradient which inhibits the flows from the contraction region (f) to the outward region (a)
leads to the vanishment of the separation vortex in the outward region (a) at low $\mathrm{Wi}$.

The Weissenberg number between the transition and elastic regions at $\mathrm{Re}=0$ in Fig.\ \ref{fig:phase}
nearly corresponds to the Weissenberg number at which the separation vortex appears.
The flows at $\mathrm{Re}=0$ are similar to those at moderate $\mathrm{Re}$
for large $\mathrm{Wi}$ as shown in Figs.\ \ref{fig:streamline200}(c) and \ref{fig:streamline0}(b).
This is because the elastic force and the pressure gradient are dominant for large $\mathrm{Wi}$ at every $\mathrm{Re}$.

\begin{figure}
  \centerline{\includegraphics[height=200pt]{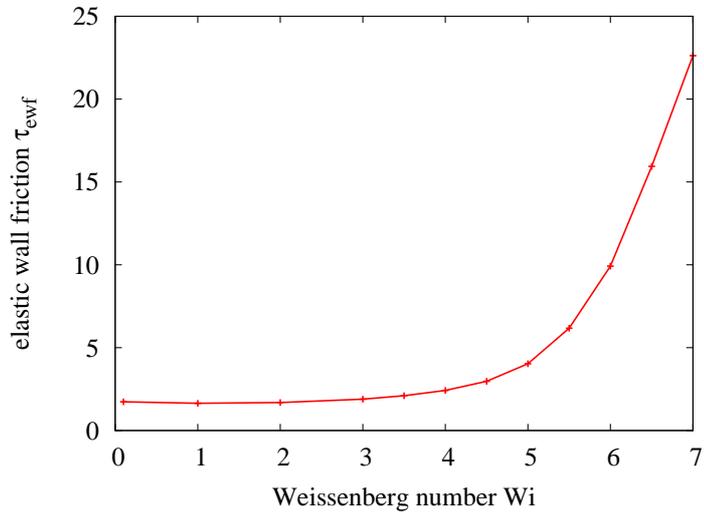}}
  \caption{(Color online) Elastic wall friction $\tau_{ewf}$ as a function of $\mathrm{Wi}$ at $\mathrm{Re}=0$. }
  \label{fig:frict0}
\end{figure}

Figure\ \ref{fig:frict0} shows the elastic wall friction as a function of $\mathrm{Wi}$.
The sudden growth of the elastic wall friction is not observed
because the flow pattern transition between the convective and transition regions does not occur.
However, the growth rates of the elastic wall frictions at the low- and high- $\mathrm{Wi}$ regions are much different.
The Weissenberg number at which the growth rate changes nearly corresponds to the Weissenberg number at which the separation vortex emerges.

\subsection{Poiseuille-type flows}

In this subsection, we show the steady solutions of Poiseuille-type flows driven by a body force
in order to elucidate the dependence of the flow pattern transition on driving forces.

\begin{figure}
  \centerline{\includegraphics[height=200pt]{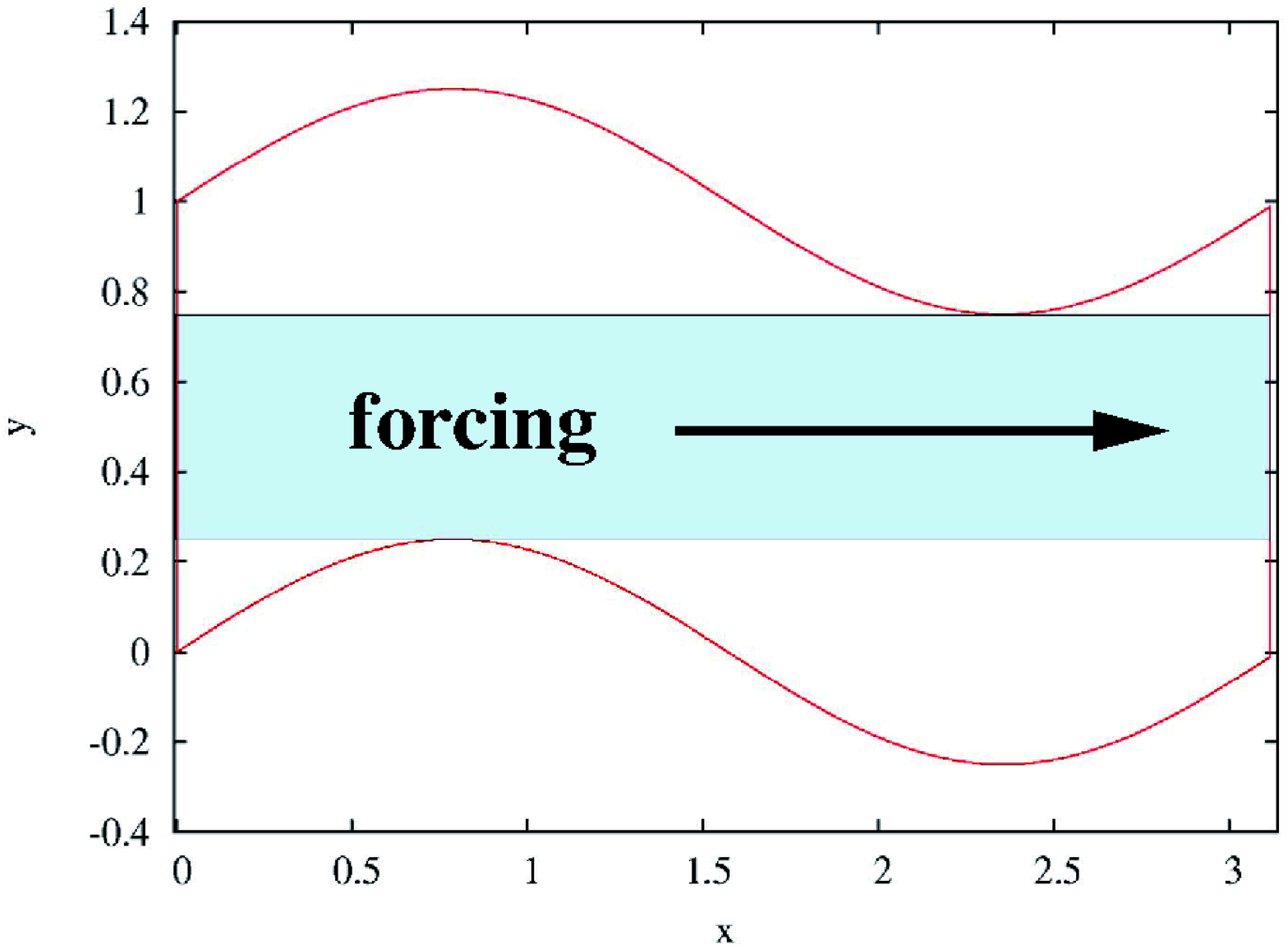}}
  \caption{(Color online) Regions in which a constant body force is imposed in Poiseuille-type flow cases.}
  \label{fig:bodyforce}
\end{figure}

We add a constant body force (constant pressure gradient) only in the rectangular regions connecting the inflow and outflow boundaries.
No body force is added in the two outward regions (a) and (e) (see Figs.\ \ref{fig:geometry}(a) and \ref{fig:bodyforce}).
The wall amplitude $A$ needs to be less than the half width of the channel $A=0.5$ in this method.
We adopt $A=0.3$ in the Poiseuille-type flow cases instead of $A=0.5$ used in the Couette-type flow cases.
The magnitude of the body force is determined under the condition that the flow rate $FR$ is fixed at every $\mathrm{Wi}$.

\begin{figure}
  \centerline{\includegraphics[height=400pt]{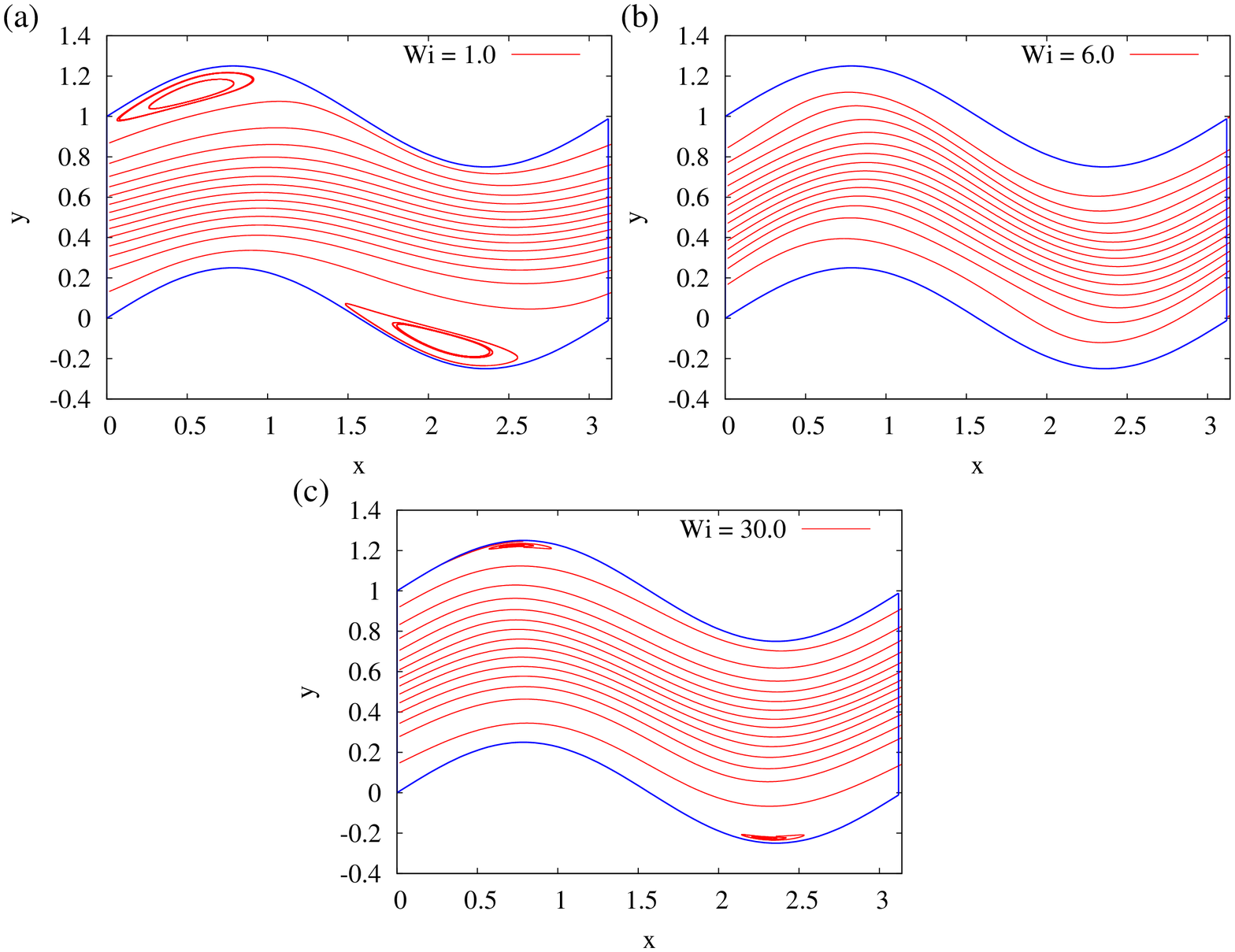} }
  \caption{(Color online) Streamlines of the Poiseuille-type flows at the low Weissenberg number $\mathrm{Wi}=1.0$,
   the moderate Weissenberg number $\mathrm{Wi}=6.0$, and the high Weissenberg number $\mathrm{Wi}=30.0$ at $\mathrm{Re}=200$.}
  \label{fig:pstreamline200}
\end{figure}

Figure\ \ref{fig:pstreamline200} shows streamlines of the steady solutions of the Poiseuille-type flows at $\mathrm{Re}=200$.
The two separation vortices appear in the outward regions (a) and (e) at the low Weissenberg number $\mathrm{Wi}=1.0$.
These separation vortices disappear at the moderate Weissenberg number $\mathrm{Wi}=6.0$,
but reappear at the high Weissenberg number $\mathrm{Wi}=30.0$.
This shows that the flow pattern transition is not influenced by the difference of the driving forces between the Couette-type and Poiseuille-type flows.

\begin{figure}
  \centerline{\includegraphics[height=200pt]{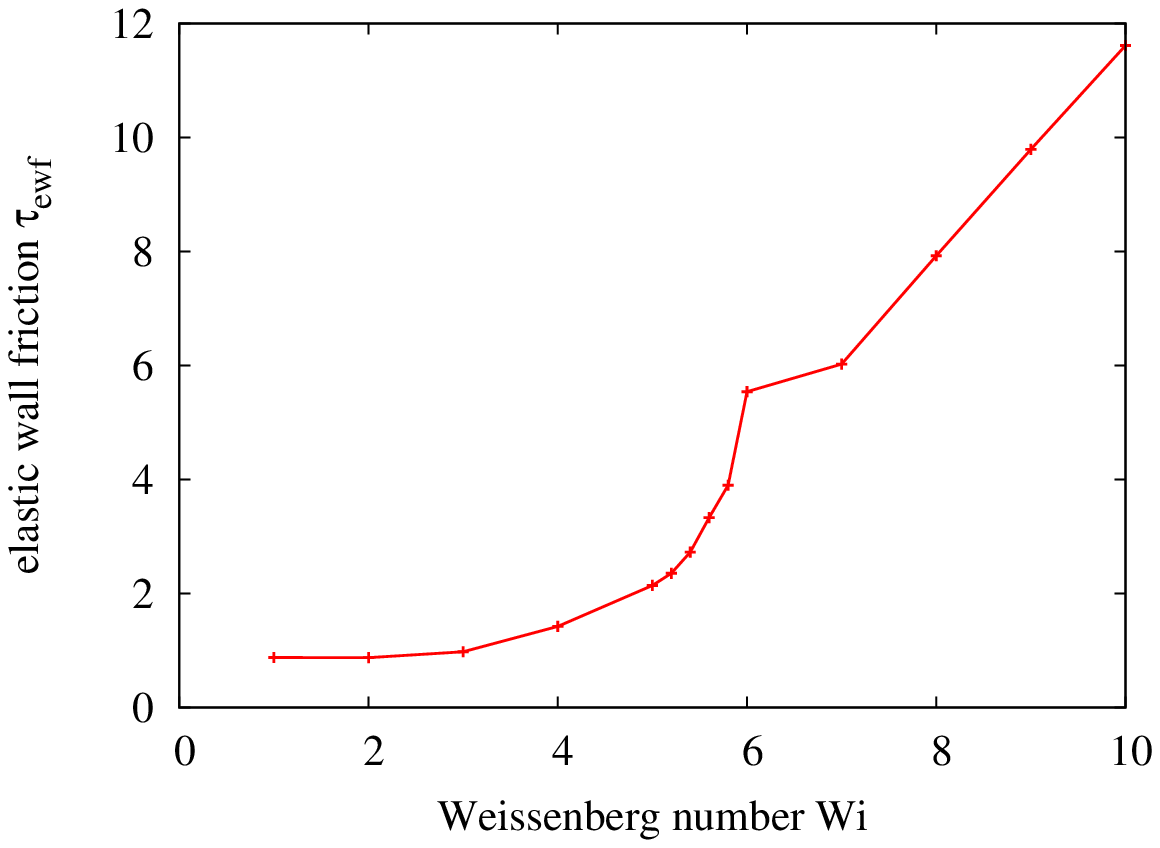}}
  \caption{(Color online) Elastic wall friction $\tau_{ewf}$ in the Poiseuille-type flows
     as a function of $\mathrm{Wi}$. }
  \label{fig:pfrict}
\end{figure}

Next, we confirm the relationship between the flow pattern transition and the variation of the elastic wall friction in the Poiseuille-type flows.
In Fig.\ \ref{fig:pfrict}, we show the elastic wall friction as a function of $\mathrm{Wi}$.
When the separation vortices disappear, the growth rate of the elastic wall friction suddenly varies at $\mathrm{Wi}=6.0$.
This result is also qualitatively consistent with that of the Couette-type flows.

Although the Poiseuille-type and Couette-type flows are qualitatively similar,
the Weissenberg numbers at which the separation vortices disappear or emerge are much different between the two flows.
The difference may result from that of the relationship between the centrifugal force and the elastic force.
In Poiseuille-type flows, the centrifugal force is larger in a bulk region than near walls and the elastic force is larger near walls than in the bulk region.
In contrast, the centrifugal force is larger near the bottom wall than in a bulk region
and the elastic force is large in the inward regions (b), (d), in the contraction regions (c), and (f) in Couette-type flows.

\section{Discussion and Concluding remarks}
\label{sec:vdiscussion}

In this study, we seek base steady solutions of the curvilinear shear flows in the wavy-walled channel.
However, the transient behavior of the relaxation process to the steady solutions is also important
when we consider the relation of the base solutions to elastic instability.
In this section, we first discuss the transient behavior using relaxation times against small perturbations on the base steady solutions
and next summarize our study.

\begin{figure}
  \centerline{\includegraphics[height=190pt]{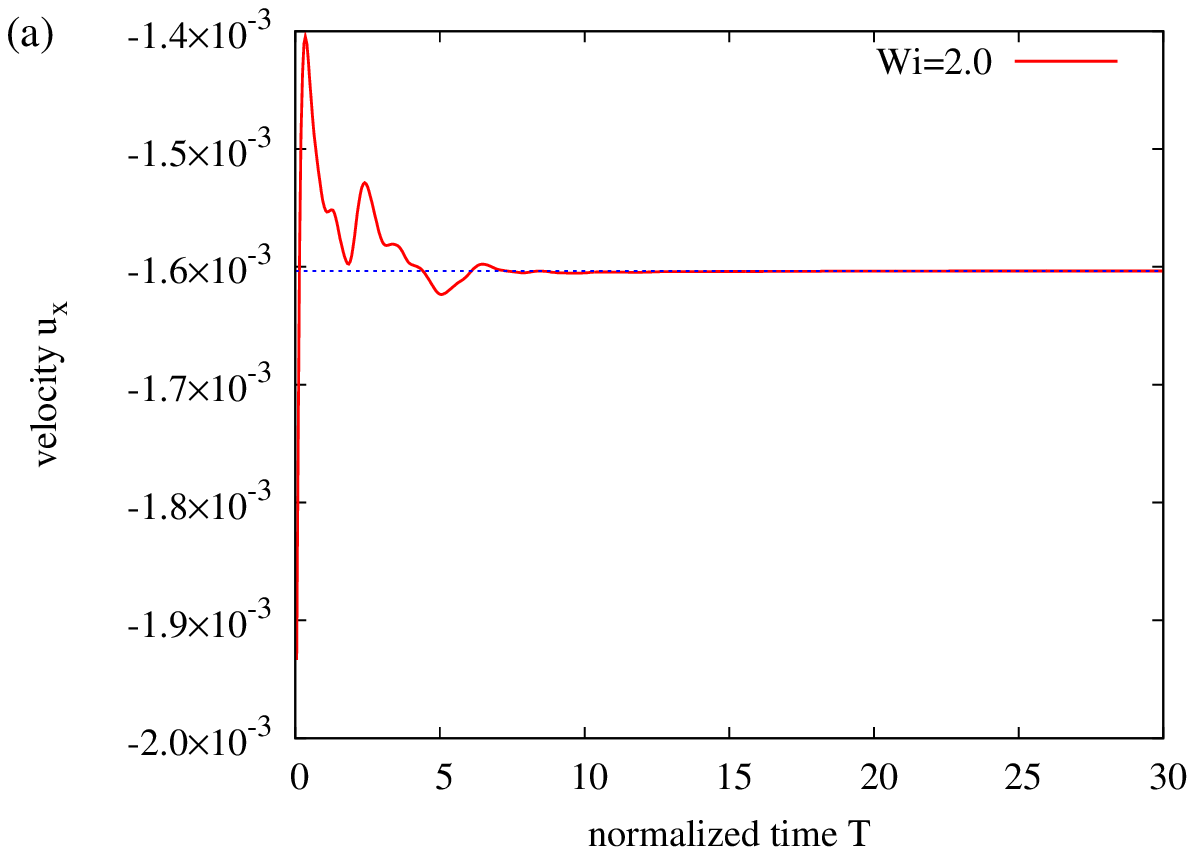}
  \includegraphics[height=190pt]{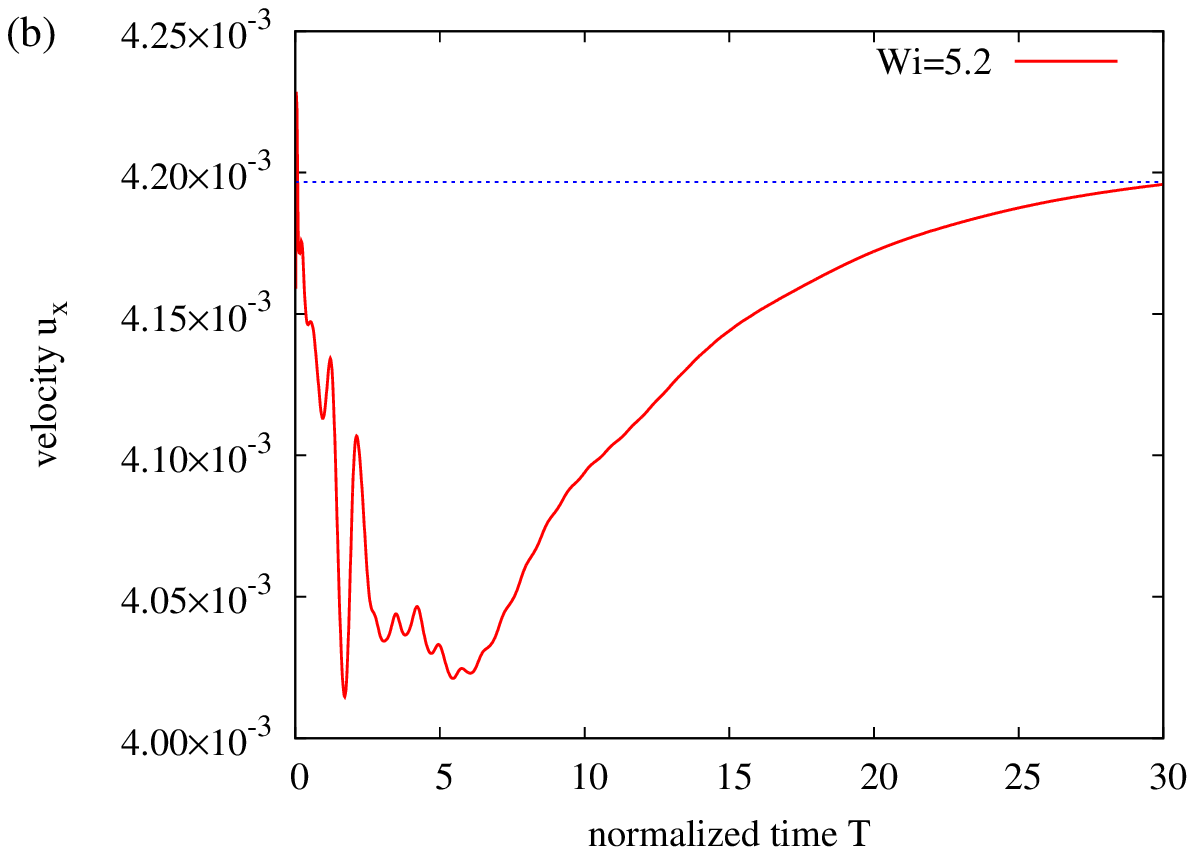}}
  \centerline{\includegraphics[height=190pt]{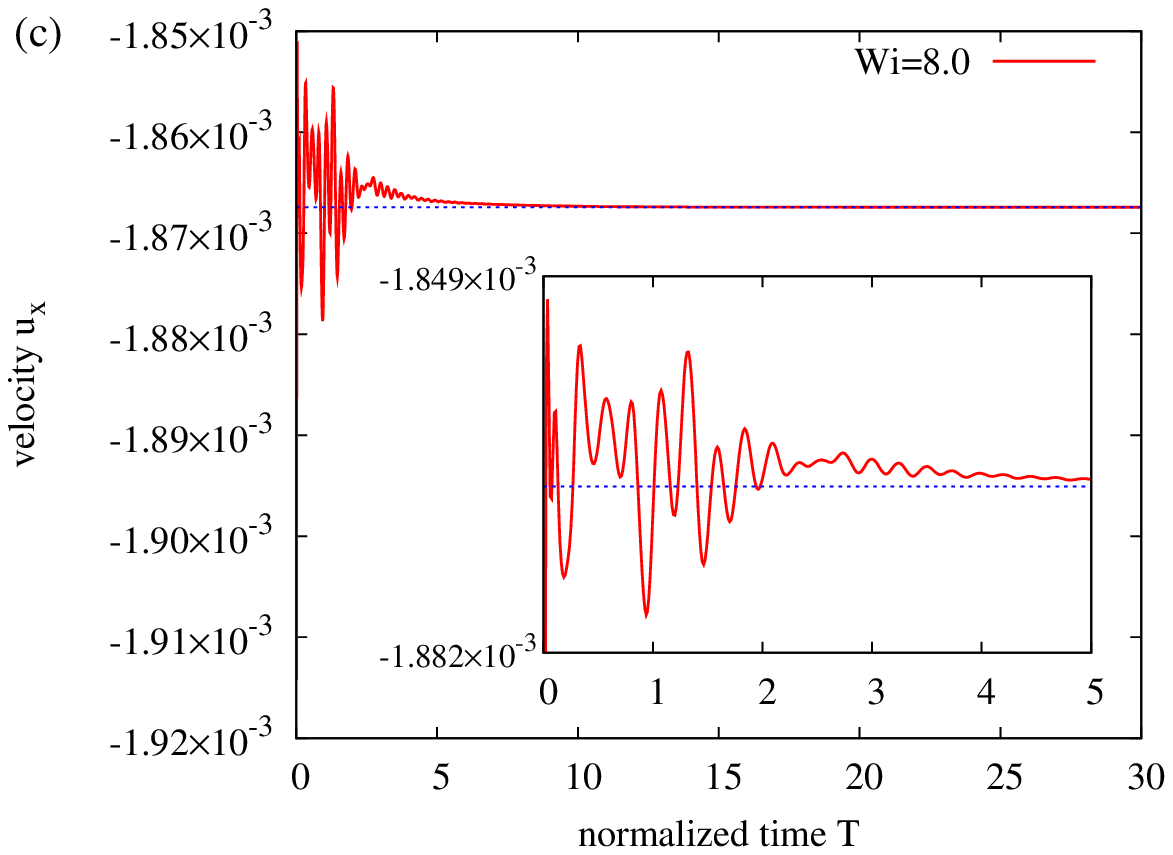}}
  \caption{(Color online) Relaxation processes of the velocity component $u_x$ against small random perturbations in (a) the convective region $\mathrm{Wi}=2.0$,
  (b) the transition region $\mathrm{Wi}=5.2$, and (c) the elastic region $\mathrm{Wi}=8.0$.
  The inset of (c) shows the relaxation process magnified in the initial short period at $\mathrm{Wi}=8.0$.
  Normalized time $T$ is defined as $T=t/\mathrm{Wi}$ at each $\mathrm{Wi}$.
  Horizontal lines denote the velocity component $u_x$ of the initial conditions.  }
  \label{fig:relaxtime}
\end{figure}

In Fig.\ \ref{fig:relaxtime}, we show the relaxation processes of the velocity component $u_x$ against small random perturbations
on the conformation tensor of the base steady solutions in the convective region $\mathrm{Wi}=2.0$,
the transition region $\mathrm{Wi}=5.2$, and the elastic region $\mathrm{Wi}=8.0$ at $\mathrm{Re}=200$.
The same perturbation is added to the three steady solutions.
The time is normalized with $\mathrm{Wi}$ which represents the relaxation time of the conformation tensor (see Eqs.\ (\ref{eq:nfenep}) and (\ref{eq:weissen})).
This figure shows that the perturbation more rapidly relaxes in the convective and elastic regions than in the transition region.
Note that the perturbation amplitude of the velocity is the largest in the transition region despite the same amplitude of the stress perturbations.
The large velocity perturbations shows that the feedback from the elastic stress to the velocity is the largest in the transition region.
We expect that the pressure gradient plays a key role in the difference of the relaxation processes between the three regions.
In the convective and elastic regions, the pressure gradient nearly accords with the centrifugal force and the elastic force, respectively.
In contrast, the pressure gradient in the transition region relatively decreases compared with the other regions (see Fig.\ \ref{fig:balance200}).
The pressure reduction and the long relaxation time in the transition region
may indicate that the pressure gradient works to depress the perturbation of the extra-stress in our flow system.
Note that the most oscillatory behavior with the smallest period is observed in the elastic region as shown in the inset of Fig.\ \ref{fig:relaxtime}(c).
The oscillatory nature in the high-$\mathrm{Wi}$ region is also observed in the earlier two-dimensional calculations of
viscoelastic extensional flows \cite{Thomases:2009}.

In order to sustain the disturbance against the relaxation due to the pressure gradient, several methods can be considered.
First, utilizing three-dimensional curved channels may lead to elastic instability
because the pressure hardly suppress spanwise secondary flows for the lack of the degree of freedom.
Spanwise secondary flows, for example G\"ortler vortices, are actually observed in three-dimensional curved-wall flows \cite{Saric:1994}.
A second normal stress difference in three-dimensional systems also influences flow behavior of curvilinear flows of viscoelastic fluids \cite{Fan:2001}
though it can work as both stabilizing and destabilizing effects \cite{Larson2:1992}.
Second, subcritical instabilities may occur against well-selected finite-amplitude perturbations in wavy-walled channels 
as in the earlier studies of plane-Poiseuille flows \cite{Atalik:2002,Morozov:2007}.
The well-selected initial conditions are generally produced using the most unstable modes derived in linear stability analyses \cite{Meulenbroek:2004}.
The initial conditions probably have the self-enhancing characteristic that the elastic effects store energy in high shear regions and dump it in less sheard regions
in order to sustain perturbations \cite{Morozov:2007}.

Finally, we summarize our study.
We have numerically investigated laminar steady flows of viscoelastic fluids
in canonical wavy-walled channels at moderate $\mathrm{Re}$ and $\mathrm{Wi}$ with spectral element method.
The steady solutions of laminar flows are classified into three groups.
The three groups correspond to three regions in the $\mathrm{Re}$-$\mathrm{Wi}$ parameter space.
The steady solutions in the three regions have remarkably different flow patterns
characterized by jet flows near the walls and the vortex detached from a bulk flow called separation vortex.
The separation vortex and the jet flow near the bottom wall emerge at low $\mathrm{Wi}$.
The vortex is temporarily vanished or becomes very small at moderate $\mathrm{Wi}$, while the vortex reappears at large $\mathrm{Wi}$.
The jet flow near the bottom wall disappears and another jet flow appears near the upper wall at moderate $\mathrm{Wi}$,
while the jet flow near the upper wall gradually reaches to the bottom wall with $\mathrm{Wi}$.
The separation vortex has been observed in numerical simulations and experiments
of both Newtonian and non-Newtonian fluid flows in wavy-walled channels \cite{Asako:1988,Arora:2002}.
The results we have obtained are consistent with those of the earlier studies
in the point that both the convective and elastic terms influence on the flow pattern formation \cite{Abu-Ramadan:2006}.
 
When the flow pattern changes, the viscous wall friction drastically decreases.
The drag reduction of laminar flows also occurs in experiments and numerical simulations of viscoelastic fluids in curved pipes \cite{Fan:2001,Jones:1976}.
The flow pattern transition also accompanies the sudden growth of the elastic wall friction.
The sudden growth of the elastic wall friction is caused by the influence of the elastic stress augmented in the bulk flow on the walls.
The growth mechanism of the elastic wall friction has similarity with that of the wall friction in Newtonian turbulent flows
in which the Reynolds stress ${\tau}^R_{ij}=\overline{\check{u}_i\check{u}_j}$ intensified near walls influences on the walls
where $\check{u}_i$ is a fluctuation component of velocity ${\bm u}=\bar{\bm u}+\check{\bm u}$.
The similarity between laminar non-Newtonian fluid flows and Newtonian turbulent flows has been long pointed out \cite{Speziale:1991}.
The similarity has been also interested in the recent studies in the context of elastic turbulence \cite{Groisman:2001,Berti:2008}.

We qualitatively described the mechanism of the flow pattern transition
by utilizing the local force relation among the centrifugal force, the elastic force, and the pressure gradient
on the basis of the variation of the terms in the momentum equation.
Through the all regions, the streamwise elastic force as well as the hoop stress is the origin of the flow pattern formations.
When a convective effect and an elastic effect balance, the abrupt pattern transition occurs.

By the scaling analysis, we shown that the stress component is proportional to the Reynolds number
on the boundary of the first transition in the $\mathrm{Re}$-$\mathrm{Wi}$ space.
This scaling coincides well with the numerical result.
The critical Reynolds number $\mathrm{Re}_c\sim70$ obtained in the scaling analysis is reasonable
as a characteristic value at which an inertial effect becomes considerable \cite{Groisman2:2004}.

We perform the computations of purely elastic flows at $\mathrm{Re}=0$.
Due to the vanishment of the centrifugal force, the separation vortex disappears at low $\mathrm{Wi}$.
The separation vortex emerges at moderate $\mathrm{Wi}$ as in the case of the moderate-$\mathrm{Re}$ calculations.
The vortex vanishment at low $\mathrm{Re}$ has been indeed observed in the computations of Newtonian fluid flows \cite{Asako:1988}.
Note that the Weissenberg number at which the separation vortex emerges is close to the Weissenberg number at which elastic turbulence
starts in the experiments of viscoelastic fluid flows for very small Reynolds number $\mathrm{Re}\ll1$ in curved-wall channels \cite{Groisman:2001}.
The modulation of the separation vortex may be related to elastic instability as observed in the earlier experiments \cite{Arora:2002}.

We finally perform the computations of Poiseuille-type flows driven by a constant body force.
The transition of the flow pattern and the growth rate of the elastic wall friction are
also observed in the Poiseuille-type flows as in the Couette-type flows.

In the future work, we plan to perform nonlinear stability analyses by utilizing the steady solutions which we have obtained in this study
in order to investigate a detailed bifurcation diagram in two-dimensional wavy-walled channels.
In addition, we will employ the three-dimensional direct numerical simulations
to confirm elastic instabilities in the three-dimensional viscoelastic wavy-walled systems.

\textbf{Acknowledgement:}
We gratefully acknowledge Sadayoshi Toh for essential discussions.
Takeshi Matsumoto, Shunsuke Kohno, and all members of our laboratory provided us with much help, advice, and comments
in the preparation of the manuscript.
This work was supported by the Grant-in-Aid for the Global COE Program
"The Next Generation of Physics, Spun from Universality and Emergence"
from the Ministry of Education, Culture, Sports, Science and Technology (MEXT) of Japan.
The numerical calculations were carried out on SX8 and Altix at YITP in Kyoto University and SX-8 at CMC in Osaka University.


\end{document}